\documentclass[twocolumn,english,aps,prb,superscriptaddress]{revtex4}
\usepackage[T1]{fontenc}
\usepackage[latin9]{inputenc}
\usepackage{amsmath}
\usepackage{amssymb}
\usepackage{graphicx}
\usepackage{esint}

\makeatletter
\@ifundefined{textcolor}{}
{%
 \definecolor{BLACK}{gray}{0}
 \definecolor{WHITE}{gray}{1}
 \definecolor{RED}{rgb}{1,0,0}
 \definecolor{GREEN}{rgb}{0,1,0}
 \definecolor{BLUE}{rgb}{0,0,1}
 \definecolor{CYAN}{cmyk}{1,0,0,0}
 \definecolor{MAGENTA}{cmyk}{0,1,0,0}
 \definecolor{YELLOW}{cmyk}{0,0,1,0}
}

\usepackage{babel}

\makeatother

\usepackage{babel}
\begin{document}

\title{Transport via double constrictions in integer and fractional topological
insulators }

\author{Chia-Wei Huang}

\affiliation{Department of Physics, Bar-Ilan University, Ramat Gan, 52900, Israel}

\author{Sam T. Carr}

\affiliation{School of Physical Sciences, University of Kent, Canterbury CT2 7NH,
UK}

\affiliation{Institut f�r Theorie der Kondensierten Materie and DFG Center for
Functional Nanostructures, Karlsruher Institut f�r Technologie, 76128
Karlsruhe, Germany}

\author{Dmitri Gutman}

\affiliation{Department of Physics, Bar-Ilan University, Ramat Gan, 52900, Israel}

\author{Efrat Shimshoni}

\affiliation{Department of Physics, Bar-Ilan University, Ramat Gan, 52900, Israel}

\author{Alexander D. Mirlin}

\affiliation{Institut f�r Theorie der Kondensierten Materie and DFG Center for
Functional Nanostructures, Karlsruher Institut f�r Technologie, 76128
Karlsruhe, Germany}

\affiliation{Institut f�r Nanotechnologie, Karlsruher Institut f�r Technologie,
76021 Karlsruhe, Germany}

\affiliation{Petersburg Nuclear Physics Institute, 188300 St.~Petersburg, Russia}

\date{\today}
\begin{abstract}
We study transport properties of the helical edge states of two-dimensional
integer and fractional topological insulators via double constrictions.
Such constrictions couple the upper and lower edges of the sample,
and can be made and tuned by adding side gates to the system. Using
renormalization group and duality mapping, we analyze phase diagrams
and transport properties in each of these cases. Most interesting
is the case of two constrictions tuned to resonance, where we obtain
Kondo behavior, with a tunable Kondo temperature. Moving away from
resonance gives the possibility of a metal-insulator transition at
some finite detuning. For integer topological insulators, this physics
is predicted to occur for realistic interaction strengths and gives
a conductance $G$ with two temperature $T$ scales where the sign
of $dG/dT$ changes; one being related to the Kondo temperature while
the other is related to the detuning. 
\end{abstract}
\maketitle

\section{Introduction}

The two-dimensional (2D) electron gas supports an amazingly broad
variety of phenomena and states. When subject to a strong magnetic
field, it gives rise to the quantum Hall effect with either integer
or fractional filling factors, depending on the strength of the Coulomb
interactions.\cite{Prange1990,Sarma1997} Such quantum Hall insulators
which have energy gaps in the bulk and gapless chiral states on the
edges are typical examples of 2D topological insulators (TIs). In
this case the presence of the magnetic fields break time reversal
symmetry.\cite{Hasan2010,Qi2011}

In contrast, there is another class of 2D TIs which preserve time
reversal symmetry and are realized in materials exhibiting strong
spin orbit interaction.\cite{Kane2005,Bernevig2006,Bernevig2006a,Konig2007}
For example, HgCdTe quantum well structures have been shown to be
in this new class. Since the edge states of the systems resemble two
copies of integer quantum Hall edge states with opposite spins propagating
in the different directions, they are also know as quantum spin Hall
insulators (QSHIs), with helical edge states. In analogy to the existence
of the fractional quantum Hall effect, 2D fractional QSHIs have been
theoretically predicted, but not yet experimentally realized.\cite{Levin2009,Neupert2011,Levin2012,Oreg2013}

Whether integer or fractional, the helical nature of the edge states
prohibits perturbations that respect time reversal invariance from
inducing elastic backscattering processes. Much theoretical work has
gone into understanding how inelastic scattering processes may give
rise to finite resistivity.\cite{Xu2006,Wu2006,Maciejko2009,Schmidt2012,Lezmy2012,Budich2012,Goth2013}
However, motivated by quantum Hall systems, there is an alternative
way to probe transport properties of the edge states. One may make
quantum point contacts (QPCs) or constrictions between the upper and
lower edges of the sample, i.e. by applying electrical side gates
to the systems. The constrictions act as impurities in the systems
allowing for backscattering between the same spin species. When the
constrictions are fully open, the conductance is given by its quantized
value $G=2e^{2}/h$.\cite{Buttiker1988} On the contrary, when the
constrictions are strong and pinched-off, the backscattering diminishes
the conductance, which may eventually fall to be zero.

Interactions turn these edge states into one-dimensional collective
modes,\cite{Giamarchi2003} which by combining the upper and lower
edges can be mapped to a spinful Luttinger liquid. The constriction
then becomes equivalent to the problem of an impurity in a spinful
Luttinger liquid which has been well studied.\cite{Kane1992} However,
the combination of the helical geometry and a local interaction gives
a curious relation between the Luttinger parameters in the charge
and spin sectors $g_{c}=g_{s}^{-1}\equiv g$,\cite{Teo2009,Hou2009,Strom2009,Liu2011}
which gives rise to new physics. For example, Teo and Kane\cite{Teo2009}
have shown the existence of some novel critical behavior in a point
contact in a QSHI.

\begin{figure}
\includegraphics[width=1\columnwidth]{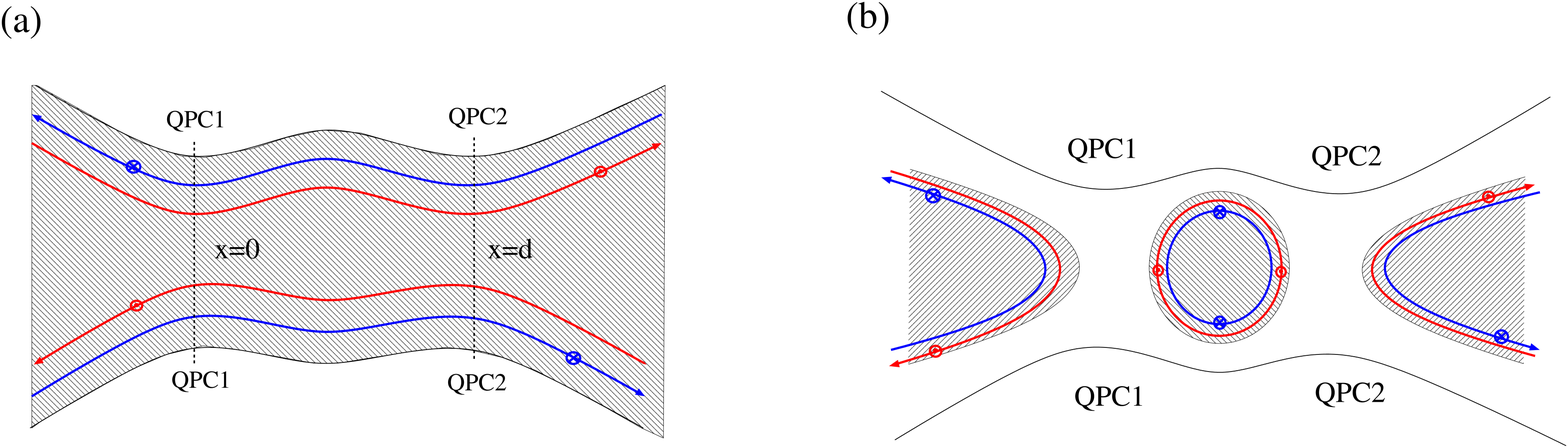} \caption{\label{fig:doubleQPC}Schematic representation of two quantum point
contacts on resonant. (a) weak backscattering limit (b) pinched-off
limit.}
\end{figure}

In addition, recent experimental progresses on transport through multiple
constrictions have allowed the study of many interesting physics problems,
such as quantum dots, Coulomb blockade, and Kondo problems.\cite{Wen1991,Kane1992,Kane1992a,Lee1992,Furusaki1993,Furusaki1993a,Maciejko2009,C.Chamon1993}
Motivated by this, we are driven to study the case of two constrictions
with a geometry as shown in Fig.~\ref{fig:doubleQPC}. Using a perturbative
renormalization group (RG) transformation,\cite{Furusaki1993,A.O.Gogolin1998,Giamarchi2003}
we calculate the conductance when the interacting electrons or quasiparticles
are weakly backscattered from the two constrictions. In the opposite
limit, when the gate voltages on the two constrictions are increased
such that the system is broken down to an island (quantum dot) in
between two leads, we use the method of instanton expansion to map
the system to its dual field,\cite{Rajaraman1982,Furusaki1993} and
calculate the corresponding RG flows in the weak link limit.

Of particular interest is the case where the double constriction is
tuned to resonance -- in this case, one obtains an emergent Kondo
effect on the island between the constrictions. Our analysis therefore
is complementary to previous discussions of Kondo impurities in topological
insulators\cite{Law2010,Seng2011,Chung2012,Eriksson2012,Posske2013,Lee2013,Chao2013,Eriksson2013}
which is based on an analysis of the Toulouse point of such models.
We also study the case where the constrictions are detuned slightly
away from resonance; which in principle can be controlled by a top
gate over the island. Over a wide parameter range, we find a metal-insulator
transition at some \textit{non-zero} value of this detuning parameter;
we will explain this metal insulator transition and show that it gives
rise to an interesting temperature dependence of the Ohmic conductance,
$G$.

We further extend the analysis to the more general case of fractional
topological insulators (FTIs), in which the filling factor is not
an integer number but a fraction with $\nu=1/m$ ($m$ is an odd integer).
In this case, the spectrum consists not only of electron like objects,
but also quasi-particles with a fractional charge of $\nu e$. Taking
quasiparticle tunneling into consideration, we make predictions about
the behavior of this exotic phase, should it be found experimentally.


Our paper is organized as follows. In section \ref{sec:Model-and-review},
we describe our model, and review the results in Refs. \onlinecite{Teo2009,Hou2009,Strom2009}.
In Sec. \ref{sec:Transmission-through-two}, we study resonant transport
properties via two constrictions in QSHIs. In Sec. \ref{sec:2d-fractional-TIs},
we discuss our model Hamiltonian and phase diagram for 2D FTIs with
a single quantum point contact, and in Sec. \ref{sec:2d-fractional-TI-2}
we generalize to the problem of a double constriction to 2D FTIs.
Finally, in Sec. \ref{sec:Conculding-remarks}, we summarize our results
and discuss possible future directions.

\section{\label{sec:Model-and-review}Model and review}

\subsection{Set-up and model Hamiltonian}

\begin{figure}
\begin{centering}
\includegraphics[width=1\columnwidth]{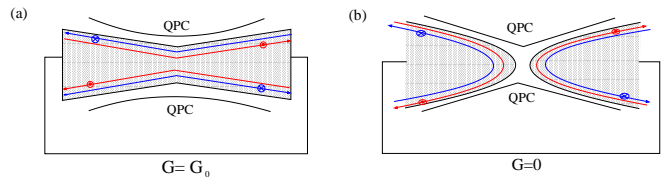} 
\par\end{centering}

\caption{\label{fig:openup}(a) Perfect transmission regime. When the quantum
point contact is open, a universal conductance $G_{0}=2\nu\frac{e^{2}}{h}$
is expected to be measured. (b) Perfect reflection regime. When the
quantum point contact is pinched off completely, the conductance is
zero (in the integer TI case, the quasiparticle is an electron and
$\nu=1$).}
\end{figure}

Before considering the double constriction geometry of Fig.~\ref{fig:doubleQPC},
we will set-up the problem and our notation by reviewing the situation
with a single constriction present. As shown in Fig. \ref{fig:openup},
the set-up is that of a two terminal Hall bar connected to a battery.
An additional gate, which creates a constriction, is added perpendicular
to the Hall bar, allowing for tunneling between the top and the bottom
edges at this point. When the constriction is weak as shown in Fig.
\ref{fig:openup}(a), the two terminal conductance is close to the
open limit of $G=2e^{2}/h$.\cite{Maslov1995,Ponomarenko1995,Safi1995}
On the contrary, when the constriction is strong, the geometry is
better represented as the pinched-off limit as shown in Fig. \ref{fig:openup}(b).
In this case, the conductance is close to zero.

The helical edge states of the sample can be understood as two copies
of integer quantum Hall systems with the two spin states of an electron
experiencing opposite effective magnetic fields. We begin with our
analysis by defining chiral boson fields $\Phi_{\eta\sigma}$, and
the density operator 
\begin{equation}
\rho_{\eta\sigma}=\frac{1}{2\pi}\partial_{x}\Phi_{\eta\sigma,}\label{eq:density}
\end{equation}
 where $\eta=R,\, L$, and $\sigma=\uparrow,\,\downarrow$. The $R\uparrow$
and $L\downarrow$ states are on the top edge of the sample, while
the $R\downarrow$ and $L\uparrow$ states are on the bottom edge.
The boson field $\Phi$ satisfies the commutation relation (we use
units $\hbar=1$, except for when we write conductance): 
\begin{equation}
\left[\Phi_{\eta\sigma}(x),\Phi_{\eta'\sigma'}(x')\right]=\pi i\eta\delta_{\eta\eta'}\delta_{\sigma\sigma'}\mathtt{sgn}(x-x').\label{eq:kacmoody}
\end{equation}
 When the short range electron interactions (with interaction strength
$\lambda_{2}$ and $\lambda_{4}$ ) are included on the edges, each
edge states can be mapped to a spinless Luttinger liquid as follows
\begin{equation}
H=H_{T}+H_{B},\label{eq:totalH}
\end{equation}
 where the Hamiltonian of the top (bottom) edge is 
\begin{eqnarray}
H^{T(B)} & = & \int\mathtt{d}x\left[\pi v_{F}\left(\rho_{R\uparrow(\downarrow)}^{2}+\rho_{L\downarrow(\uparrow)}^{2}\right)\right.\label{eq:topedge}\\
 & + & \left.\lambda_{2}\rho_{R\uparrow(\downarrow)}\rho_{L\downarrow(\uparrow)}+\lambda_{4}\left(\rho_{R\uparrow(\downarrow)}^{2}+\rho_{L\downarrow(\uparrow)}^{2}\right)\right].\nonumber 
\end{eqnarray}

By introducing new boson fields 
\begin{eqnarray}
\Phi_{R\uparrow(\downarrow)} & = & \sqrt{\frac{1}{2}}\left(\theta_{T(B)}-\phi_{T(B)}\right),\label{eq:f1}\\
\Phi_{L\downarrow(\uparrow)} & = & -\sqrt{\frac{1}{2}}\left(\theta_{T(B)}+\phi_{T(B)}\right),\label{eq:f2}
\end{eqnarray}
 we diagonalize the total Hamiltonian in Eq. (\ref{eq:totalH}) as
follows 
\begin{equation}
H=\frac{v}{4\pi}\sum_{i=T,B}\int\mathtt{d}x\left[\frac{1}{g}\left(\nabla\theta_{i}\right)^{2}+g\left(\nabla\phi_{i}\right)^{2}\right],\label{eq:second}
\end{equation}
 where the boson fields obey a new commutation relation 
\begin{equation}
\left[\phi(x),\theta(x')\right]=i\pi\mathtt{sgn}(x-x'),
\end{equation}
 and 
\begin{equation}
v=v_{F}\sqrt{\left(1+\frac{\lambda_{4}}{\pi v_{F}}\right)^{2}-\left(\frac{\lambda_{2}}{2\pi v_{F}}\right)^{2}},\label{eq:velocity}
\end{equation}
 and 
\begin{equation}
g=\sqrt{\frac{1+\lambda_{4}/\pi v_{F}-\lambda_{2}/2\pi v_{F}}{1+\lambda_{4}/\pi v_{F}+\lambda_{2}/2\pi v_{F}}}.\label{eq:g}
\end{equation}

Since constrictions couple top and bottom edges, it is instructive
to work with a new basis which is a linear combination of them. This
maps the two spinless Luttinger liquids onto a single spinful Luttinger
liquid with physical quantities spin $(s)$ and charge $(c)$ density.
To account for the distribution of spin between the two edges, the
transformation required is: \begin{subequations}\label{eq:sall}
\begin{eqnarray}
(\theta_{T}+\theta_{B}) & = & \begin{array}{c}
\sqrt{2}\theta_{c},\end{array}\label{eq:s1}\\
(-\theta_{T}+\theta_{B}) & = & \begin{array}{c}
\sqrt{2}\phi_{s},\end{array}\label{eq:s2}\\
(\phi_{T}+\phi_{B}) & = & \begin{array}{c}
\sqrt{2}\phi_{c},\end{array}\label{eq:s3}\\
(-\phi_{T}+\phi) & = & \begin{array}{c}
\sqrt{2}\theta_{s}.\end{array}\label{eq:s4}
\end{eqnarray}
 \end{subequations} Inserting Eqs. (\ref{eq:sall}) into Eq. (\ref{eq:second}),
we obtain a new Hamiltonian with charge (with subscript $c$) and
spin (with subscript $s$) sectors as follows: 
\begin{equation}
H=\frac{v}{4\pi}\sum_{a=c,s}\int\mathtt{d}x\left[\frac{1}{g_{a}}\left(\nabla\theta_{a}\right)^{2}+g_{a}\left(\nabla\phi_{a}\right)^{2}\right],\label{eq:fourth}
\end{equation}
 where 
\begin{equation}
g_{c}=\frac{1}{g_{s}}=g.
\end{equation}

The general expression for the original bosonic fields in Eq. (\ref{eq:density})
in terms of $\theta_{c}$, $\theta_{s}$, $\phi_{c}$ and $\phi_{s}$
is 
\begin{equation}
\Phi_{\eta\sigma}=\frac{1}{2}\left[\left(\eta\theta_{c}-\phi_{c}\right)+\sigma\left(\eta\theta_{s}-\phi_{s}\right)\right],\label{eq:veryfield}
\end{equation}
 where $\eta=R,\, L=+,-$, and $\sigma=\uparrow,\downarrow=+,-$.
In order to complete our bosonized representation of the problem,
we must also give the relation between the bosonized fields and the
original Fermionic operators: 
\begin{equation}
\Psi_{\eta\sigma}=\frac{1}{\sqrt{2\pi\alpha}}e^{i\Phi_{\eta\sigma}}
\end{equation}
 where $\alpha$ is a short distance cutoff of the field theory. In
what follows, we use this notation and the spinful Luttinger Hamiltonian
defined in Eq. (\ref{eq:fourth}) to carry out the calculations in
this report. Unless otherwise stated, we also use the convention that
the Fermi velocity $v_{F}=1$.

It is also worth making a comment at this point about our assumption
of the conservation of $S_{z}$. In general, two dimensional topological
insulators occur in materials with strong spin-orbit coupling, meaning
that $S_{z}$ is not conserved. This is crucial when looking at scattering
mechanisms on a single edge,\cite{Xu2006,Wu2006,Maciejko2009,Schmidt2012,Lezmy2012,Budich2012}
where one must violate conservation of $S_{z}$ in order to get any
scattering at all. However, when a constriction is present as we will
consider in this work, the dominant scattering mechanism is through
the constriction and does not require $S_{z}$ non-conserving terms.\cite{Teo2009,Eriksson2012,Eriksson2013,Lee2013}
We will therefore not consider this in the current paper.

\subsection{Review: single constriction in quantum spin Hall insulator }

In this subsection we review the renormalization group flows in the
two limits as shown in Fig. \ref{fig:openup}.\cite{Teo2009,Strom2009,Hou2009}
This will form the starting point for extending the results to two
constrictions and/or the FTIs. It will also be important to summarize
the results here when we later look at off-resonance tunneling through
the double constriction in section \ref{sec:off-resonance}.

\subsubsection{Weak backscattering limit in quantum spin Hall fluids}

\begin{figure}
\includegraphics[width=1\columnwidth]{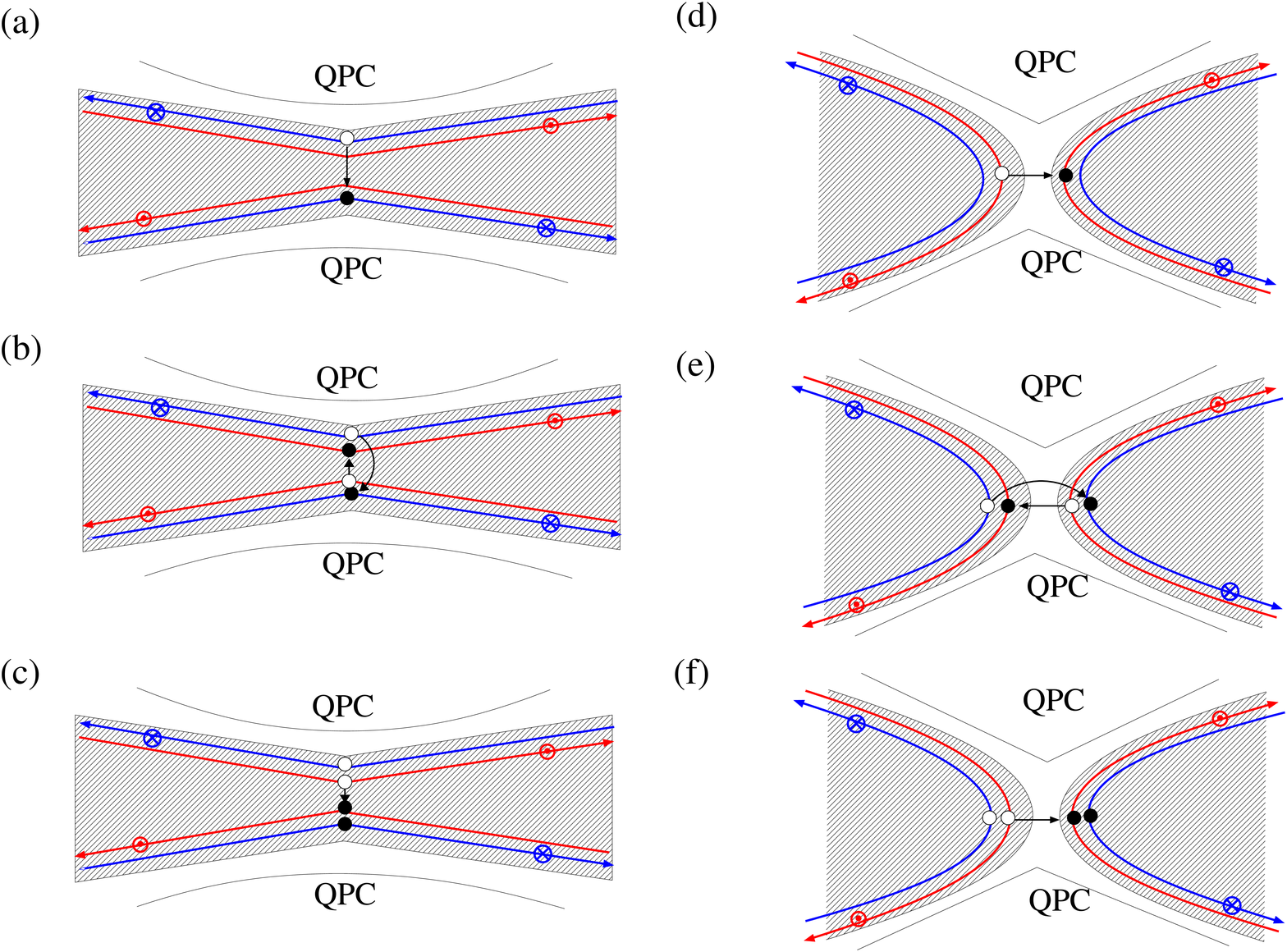} \caption{\label{fig:All}(a-c) Schematic representation of backscattering processes
in a near perfectly transmitting phase ($C_{c}C_{s}$). (a) describes
the single quasiparticle of charge $\nu e$ tunneling process with
perturbation operator $\upsilon_{e}^{q}$ (in the integer TI case,
the quasiparticle is an electron and $\nu=1$). When spin is conserved,
the particle can either tunnel from $L\uparrow$ to $R\uparrow$ or
from $L\downarrow$ to $R\downarrow$. (b) describes a particle pair
tunneling process with opposite spins by an operator $\upsilon_{c}^{q}$.
(c) describes the transfer of 2 quasiparticles with charge $2\nu e$
from one edge to the other by an operator $\upsilon_{s}^{q}$. (d-f)
Schematic representation of tunneling processes in a nearly perfect
reflecting phase ($I_{c}I_{s}$). (d) describes single electron tunneling
process with perturbation operator $t_{e}$. When spin is conserved,
the $\mathit{electron}$ either tunnels from $L\uparrow$ to $R\uparrow$
or from $L\downarrow$ to $R\downarrow$. (e) describes an electron
pair tunneling process with opposite spins by a operator $t_{c}$.
(f) describes the charge transfer of 2 electrons from one side to
the other by an operator $t_{s}$ (in the integer TI case, the quasiparticle
is an electron and $\nu=1$).}
\end{figure}

When the QPC is fully open, a perfect conductance $G=2\frac{e^{2}}{h}$
is expected to be observed.\cite{Maslov1995,Ponomarenko1995,Safi1995}
However in general, the QPC acts as a local impurity which gives rise
to backscattering processes, as shown in Fig. \ref{fig:All}. Single
backscattering processes as depicted in Fig. \ref{fig:All}(a) are
always present. However in the presence of interactions, it is also
important to consider coherent two-particle backscattering processes
illustrated in Figs.~\ref{fig:All}(b) and \ref{fig:All}(c).\cite{Teo2009}
Technically, one may think of these terms as generated by RG transformation.

Taking all these terms into account, the backscattering Hamiltonian
is therefore 
\begin{eqnarray}
H_{b} & = & \upsilon_{_{e}}\left(\Psi_{R\uparrow}^{\dagger}\Psi_{L\uparrow}+\Psi_{R\downarrow}^{\dagger}\Psi_{L\downarrow}+h.c.\right)\label{eq:backscatter}\\
 &  & +2\pi\alpha\upsilon_{_{c}}\left(\Psi_{R\uparrow}^{\dagger}\Psi_{R\downarrow}^{\dagger}\Psi_{L\uparrow}\Psi_{L\downarrow}+h.c.\right)\nonumber \\
 &  & +2\pi\alpha\upsilon_{_{s}}\left(\Psi_{R\uparrow}^{\dagger}\Psi_{L\downarrow}^{\dagger}\Psi_{L\uparrow}\Psi_{R\downarrow}+h.c.\right),\nonumber \\
 & = & \frac{\upsilon_{_{e}}}{2\pi\alpha}\cos\theta_{c}\cos\theta_{s}+\frac{\upsilon_{_{c}}}{2\pi\alpha}\cos2\theta_{c}+\frac{\upsilon_{_{s}}}{2\pi\alpha}\cos2\theta_{s}\nonumber 
\end{eqnarray}
 Here $\upsilon_{_{e}}$ stands for the $\mathit{single}$ electron
backscattering process across the QPC, $\upsilon_{_{c}}$ represents
a pair backscattering with opposite spins, and $\upsilon_{_{s}}$
represents the transfer of 2$e$ charged particles from the top to
the bottom edges. The normalization is chosen so as to make all these
parameters dimensionless.

The leading order RG flows for each process is as follows: 
\begin{equation}
\frac{d\upsilon_{_{a}}}{dl}=\left(1-\Delta_{\upsilon a}\right)\upsilon_{_{a}},\label{eq:generalRG}
\end{equation}
 where $l=\ln\Lambda/T$, and the scaling dimensions are given by:
\begin{subequations}\label{eq:vall} 
\begin{eqnarray}
\Delta_{\upsilon e} & = & \left(g+g^{-1}\right)/2,\\
\Delta_{\upsilon c} & = & 2g,\label{eq:vc}\\
\Delta_{\upsilon s} & = & 2g^{-1}.\label{eq:vs}
\end{eqnarray}
 \end{subequations} These equations show that while $\Delta_{\upsilon e}$
is always greater than or equal to one and the single particle backscattering
is always irrelevant, the pair backscattering processes may or may
not be relevant depending on the value of $g$. For $1/2<g<2$, which
includes the case of weak interactions $g\approx1$, all backscattering
processes are irrelevant, and the conducting phase is a stable fixed
point. The RG flows for the weak backscattering limit for the full
range of interaction $g$ is plotted in the upper part of Fig. 6(a)
in Ref. \onlinecite{Teo2009}.

\subsubsection{Weak tunneling limit in quantum spin Hall insulator }

When the QPC is completely pinched off, the conductance is expected
to be $G=0$ {[}see geometry in Fig.~\ref{fig:All}(d) - (f){]}.
In the vicinity of this point, one allows for weak tunneling between
the two halves of the sample. Taking into account both single and
two particle processes, as illustrated in Fig. \ref{fig:All}(d) -
(f), the model Hamiltonian is described as follows: 
\begin{eqnarray}
H_{t} & = & t_{_{e}}\left(\Psi_{+\uparrow}^{\dagger}\Psi_{-\uparrow}+\Psi_{+\downarrow}^{\dagger}\Psi_{-\downarrow}+h.c.\right)_{x=0}\label{eq:Ht1}\\
 &  & +2\pi\alpha t_{_{c}}\left(\Psi_{+\uparrow}^{\dagger}\Psi_{+\downarrow}^{\dagger}\Psi_{-\uparrow}\Psi_{-\downarrow}+h.c.\right)_{x=0}\nonumber \\
 &  & +2\pi\alpha t_{_{s}}\left(\Psi_{+\uparrow}^{\dagger}\Psi_{-\downarrow}^{\dagger}\Psi_{-\uparrow}\Psi_{+\downarrow}+h.c.\right)_{x=0},\nonumber \\
 & = & \frac{t_{_{e}}}{2\pi\alpha}\cos\bar{\phi}_{c}\cos\bar{\phi}_{s}+\frac{t_{_{c}}}{2\pi\alpha}\cos2\bar{\phi}_{c}+\frac{t_{_{s}}}{2\pi\alpha}\cos2\bar{\phi}_{s}.\nonumber 
\end{eqnarray}
 Here $+/-$ represents an infinitesimal displacement from the right
and left sides of the QPC located at $x=0$, $\bar{\phi}_{\alpha}=\phi_{\alpha}(+)-\phi_{\alpha}(-)$,
$t_{_{e}}$ stands for the $\mathit{single}$ $electron$ tunneling
process across the QPC, $t_{_{c}}$ represents electron pairs tunneling
with opposite spins, and $t_{_{s}}$ represents the transfer of 2$e$
electron charge from the left to the right sides. As before, the leading
order RG flows for each process is: 
\begin{equation}
\frac{dt_{_{a}}}{dl}=\left(1-\Delta_{ta}\right)t_{_{a}},
\end{equation}
 with scaling dimensions \begin{subequations}\label{eq:scaling_dimensions_t}
\begin{eqnarray}
\Delta_{te} & = & \left(g+g^{-1}\right)/2,\\
\Delta_{tc} & = & 2g^{-1},\\
\Delta_{ts} & = & 2g.
\end{eqnarray}
 \end{subequations} For $1/2<g<2$, all tunneling processes are irrelevant,
and the insulating phase is a stable fixed point. For more generic
$g$, the RG flow is plotted in the lower part of Fig. 6(a) in Ref.~\onlinecite{Teo2009}.

The fact that both the insulating and conducting fixed points are
stable for $1/2<g<2$ means that there must be an intermediate unstable
fixed point, separating the conducting and insulating phases. This
was analyzed in detail in Ref.~\onlinecite{Teo2009}, and the resulting
phase diagram is shown in Fig.~6(a) of this work. For later reference,
the phase boundary line is also indicated in Fig.~\ref{fig:RG_double_ITI}
of the present paper (see the red dashed curve in the middle of the
graph). We follow the notation of Ref.~\onlinecite{Teo2009} and
denote the insulating phase by II and the conducting one by CC. There
are also phases which are a charge conductor but spin insulator (CI)
and vice versa (IC).


\section{Transmission through two constrictions and Kondo resonance in integer
quantum spin Hall insulators}

\label{sec:Transmission-through-two}

\subsection{Renormalization group flow for the weak backscattering and weak tunneling
limits}

We now turn to the case of a double constriction as depicted in Fig.~\ref{fig:doubleQPC}.
Such a double constriction is not only technically feasible in experiments
but also contains interesting physics, for example quantum dot physics,
Coulomb blockade and Kondo physics. As shown in Fig. \ref{fig:doubleQPC},
this setup is modeled as two tunneling terms between the upper and
lower helical edges at locations $x=0$ and $x=d$. After following
the previous mapping to the spinful Luttinger liquid, this takes the
form of two impurities at these locations, each of which may give
rise to backscattering.

Using the standard technique of integrating over degrees of freedom
at points other than $x=0,d$, one obtains the local action as follows:\cite{Kane1992,Furusaki1993a,Giamarchi2003}
\begin{eqnarray}
S & = & S_{0}+\int\mathtt{d}\tau\, V_{eff},\label{eq:dualmap}
\end{eqnarray}
 where 
\begin{eqnarray}
S_{0} & = & \frac{1}{\beta}\sum_{\alpha,\omega_{n}}\frac{1}{4\pi g_{\alpha}}\left|\omega_{n}\right|\left|\theta_{\alpha}\left(\omega_{n}\right)\right|^{2},
\end{eqnarray}
 with $\alpha=+c,-c,+s,-s$, the Matsubara frequencies $\omega_{n}=2\pi nT$,
and 
\begin{eqnarray}
V_{eff} & = & V\left(\cos\frac{\theta_{+c}}{\sqrt{2}}\cos\frac{\theta_{-c}}{\sqrt{2}}\cos\frac{\theta_{+s}}{\sqrt{2}}\cos\frac{\theta_{-s}}{\sqrt{2}}\right.\nonumber \\
 &  & \left.+\sin\frac{\theta_{+c}}{\sqrt{2}}\sin\frac{\theta_{-c}}{\sqrt{2}}\sin\frac{\theta_{+s}}{\sqrt{2}}\sin\frac{\theta_{-s}}{\sqrt{2}}\right)\nonumber \\
 &  & +\frac{U_{c}}{2}\left(\theta_{-c}-\theta_{-c_{0}}\right)^{2}+\frac{U_{s}}{2}\left(\theta_{-s}-\theta_{-s_{0}}\right)^{2}.\label{eq:veffkondo}
\end{eqnarray}
 In the above expressions, $\theta_{\pm,a}=\left(\theta_{1,a}\pm\theta_{2,a}\right)/\sqrt{2}$,
the subscripts $1$ and $2$ denoting the original field operators
$\theta_{a}$ at $x=0$ and $x=d$ respectively. The $+$ sign then
corresponds to the spin or charge that has been transferred through
the junction, and the $-$ sign to the spin or charge between the
barriers (we will refer to this region as the quantum dot). The barrier
strength is $V$, while $U_{c}$ and $U_{s}$ are phenomenological
parameters introduced to describe the interactions (charging energy
and exchange energy) of the dot. Finally, $\frac{\sqrt{2}}{\pi}\theta_{-c_{0}}$
and $\frac{\sqrt{2}}{\pi}\theta_{-s_{0}}$ physically correspond to
the equilibrium values of charge and spin in the dot and may be controlled
via further external gates. We will limit ourselves to the case $\theta_{-s_{0}}=0$,
which physically means there is no external magnetic field present.

The weak constriction limit as depicted in Fig.~\ref{fig:doubleQPC}(a)
corresponds to the case when $V\ll U_{c},U_{s}$ in Eq.~(\ref{eq:veffkondo}).
One therefore first minimizes the terms involving $U$ and then treats
$V$ as a perturbation on top of this. In the particular case where
the distance between constrictions $d$ and charging gates are tuned
such that 
\begin{equation}
\theta_{-c_{0}}=\frac{\pi}{\sqrt{2}}(2n+1)\label{eq:resonance_condition}
\end{equation}
 for any integer $n$, the double constriction is on resonance and
the barrier strength $V$ in Eq.~(\ref{eq:veffkondo}) does nothing
to first order. The present physical picture therefore resembles a
single impurity problem, but with the single particle backscattering
process removed. The two particle processes however remain: $\upsilon_{c}$
backscattering a pair of electron with opposite spins, and $\upsilon_{s}$
backscattering two electrons from the top to the bottom edges. The
scaling dimensions for these two operators are identical to Eqs. (\ref{eq:vc})
and (\ref{eq:vs}).

In the strong constriction limit when $V$ is the largest energy scale
in the problem, the minimization of the $V$ term in Eq.~(\ref{eq:veffkondo})
gives us the conditions that the number of electrons $\frac{\sqrt{2}}{\pi}\theta_{-c}$
and twice the spin $\frac{\sqrt{2}}{\pi}\theta_{-s}$ on the dot are
either both even integers, or both odd integers. Further applying
the resonance condition in Eq. (\ref{eq:resonance_condition}) above
means that there are two degenerate spin states of the dot, analogous
to a Kondo problem as shown in Fig. \ref{fig:doubleQPC}(b).


Following the same line of reasoning as for the single constriction
case, we now identify the possible relevant tunneling processes that
may occur in the Kondo limit, and rewrite the problem in this dual
description. The details of this are shown in Appendix \ref{sec:KondoHamiltonian},
where following the approach of Ref.~\onlinecite{Teo2009} we obtain
the resultant tunneling Hamiltonian 
\begin{multline}
H_{t}^{K}=\frac{t_{_{e}}}{2\pi\alpha}\cos\bar{\phi}_{+c}\cos\bar{\phi}_{+s}+\frac{\tilde{t}_{_{c}}}{2\pi\alpha}\cos\bar{\phi}_{+c}\cos\bar{\phi}_{-s}\\
+\frac{\tilde{t}_{_{s}}}{2\pi\alpha}\cos\bar{\phi}_{+s}\cos\bar{\phi}_{-s}.\label{eq:kondotunneling}
\end{multline}
 Here, $\bar{\phi}_{+c}$ and $\bar{\phi}_{+s}$ are fields associated
with charge and spin transfer from the left to the right, exactly
analogous to the single impurity case in Eq.~(\ref{eq:Ht1}). The
remaining field, $\bar{\phi}_{-s}$ is associated with changes of
spin on the dot; and therefore must be treated carefully as there
are only two spin states allowed on the dot in the Kondo limit: $\pm1/2$.
More is said about this in Appendix \ref{sec:KondoHamiltonian}, where
the relationship between Eq.~(\ref{eq:kondotunneling}) and an instanton
expansion of action in Eq. (\ref{eq:dualmap}) is given.


\begin{figure}
\includegraphics[width=0.9\columnwidth]{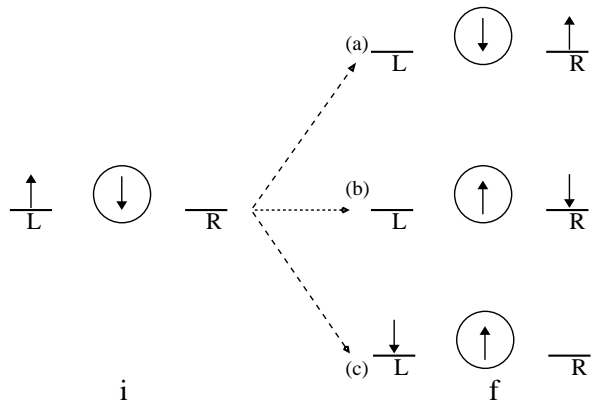} \caption{\label{fig:Resonant}Tunneling processes in the resonant tunneling
case, pictorially demonstrating the terms in Eq.~(\ref{eq:kondotunneling}).
Each process is characterized by the charge transferred from left
to right, defined as $(\Delta Q_{R}-\Delta Q_{L})/2$, and the spin
transferred through the dot defined as $(\Delta S_{R}-\Delta S_{L})/2$.
(a) Single electron tunneling process, transferring both spin and
charge across the junction; (b) Charge tunneling event, carrying a
\textit{single charge} but no spin across the junction; (c) Spin tunneling
event, carrying a \textit{single unit of spin} but no charge across
the junction. Processes (b) and (c) also involve a change of state
of the dot.}
\end{figure}

The physical meaning of the three processes in Eq.~(\ref{eq:kondotunneling})
is as follows: $t_{e}$ can be thought of a single electron tunneling
through the junction without changing the spin of the dot, $\tilde{t}_{_{c}}$
refers to a single charge transferred through the junction accompanied
by a spin flip both on the dot and the incoming electron, and $\tilde{t}$$_{_{s}}$
involves spin exchange between the dot and one of the leads. These
three processes are schematically represented in Fig.~\ref{fig:Resonant}.


The RG flow equations for this dual picture in Eq.~(\ref{eq:kondotunneling})
to second order are as follows: \begin{subequations}\label{eq:resonnatOurAll}
\begin{eqnarray}
\frac{dt_{_{e}}}{dl} & = & \left[1-\frac{1}{2}\left(g+g^{-1}\right)\right]t_{_{e}}+AgK_{s}\tilde{t}_{_{c}}\tilde{t}_{_{s}},\label{eq:resonnatOur1}\\
\frac{d\tilde{t}_{_{c}}}{dl} & = & \left[1-\frac{1}{2}\left(K_{s}g+g^{-1}\right)\right]\tilde{t}_{_{c}}+Agt_{_{e}}\tilde{t}_{_{s}},\label{eq:resonnatOur2}\\
\frac{d\tilde{t}_{_{s}}}{dl} & = & \left[1-\frac{g}{2}\left(1+K_{s}\right)\right]\tilde{t}_{_{s}}+\frac{At_{_{e}}\tilde{t}_{_{s}}}{g},\label{eq:resonnatOur3}
\end{eqnarray}
 and 
\begin{equation}
\frac{dK_{s}}{dl}=-\left(\frac{\tilde{t}_{_{c}}^{2}}{g}+g\tilde{t}_{_{s}}^{2}\right)K_{s}.\label{eq:kdependece}
\end{equation}
 \end{subequations} The parameter $K_{s}$ has the initial condition
$K_{s}(l=0)=1$, and appears as a consequence of the special considerations
of the $\bar{\phi}_{-s}$ field mentioned above.\cite{Kane1992} The
constant $A$ is non-universal; for analysis purposes we take it to
be $1$, as the results are not strongly affected by its precise value.

For small $t_{s}$ and $g\neq1$, the linear terms on the right hand
side of Eqs. (\ref{eq:resonnatOur1} - \ref{eq:resonnatOur3}) are
sufficient to describe the RG flows for the two barriers at resonant
case. However for $g\rightarrow1$, the linear terms on the right
hand side vanish, thus those quadratic terms might become important
and change the phase boundary. In fact, this will turn out not to
be the case, as we show in the next subsection.

\subsection{Phase diagram and discussion for resonant double impurity problem}

\begin{figure}
\includegraphics[width=1\columnwidth]{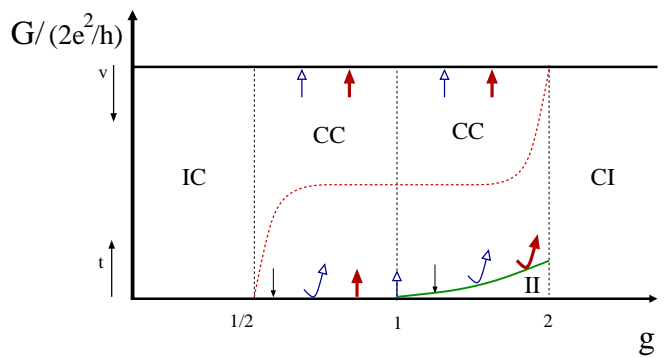} \caption{\label{fig:RG_double_ITI} 
RG flow for a resonant double barrier in an integer QSHI. The result
for the weak backscattering limit is shown on the upper part of the
figure: blue (with hollow arrow heads) and red arrows are flows for
processes transferring two charges and two spins respectively. The
result for the weak link limit is shown in the lower part of the figure,
where the thin black arrows represent single electron tunneling events,
blue arrows (with hollow arrow heads) represent \textit{single charge}
tunneling and thick red arrows represent \textit{single spin} tunneling
events. The green solid line (in the bottom) represents the phase
boundary between the conducting and insulating phases for $1<g<2$.
For comparison, the red dashed curve (in the middle of the graph)
represents the phase boundary for the case of a single constriction.\cite{Teo2009} }
\end{figure}

To begin our analysis of the phase diagram for the resonant double
constriction, we will ignore the quadratic terms in Eqs. (\ref{eq:resonnatOur1}
- \ref{eq:resonnatOur3}), or in other words we will start with the
case $A=0$. Even with this in mind, the scaling dimension analysis
is more complicated than for the single constriction case, as the
scaling dimension of the $\tilde{t}_{_{c}}$ and $\tilde{t}_{_{s}}$
operators depends on the parameter $K_{s}$, which also flows. However,
$K_{s}$ always starts at one, and from Eq. \eqref{eq:kdependece}
we see that it always flows towards zero (it may however have a limit
at some intermediate value). We can therefore look at the scaling
dimensions (and therefore the relevance) of the tunneling operators
at these two limits of $K_{s}=0,1$. This flow is summarized in Fig.~\ref{fig:RG_double_ITI}
-- the curved arrows indicate flow which is initially irrelevant,
but some time later in the flow (as $K_{s}$ changes) may become relevant.
The figure also shows the stability of the conducting (weak constriction)
phase, which is the same as the single barrier case, but with the
single electron backscattering removed.

For the weak backscattering limit, the pair backscattering terms are
irrelevant under RG for $1/2<g<2$ and we therefore predict that the
CC phase is stable in this parameter regime. In the weak link limit
however, more interesting things may happen. First, we note that the
electron tunneling term $t_{_{e}}$ is always irrelevant, and decouples
from the equations (when $A=0$), so may be ignored for the present
discussion. For $1/2<g<2$, the initial flow of the single charge
tunneling term $\tilde{t}_{_{c}}$ is towards weak coupling; however
as $K_{s}$ decreases this term may become relevant and drive the
system towards the CC phase. Now, if $1/2<g<1$, the spin tunneling
term $\tilde{t}_{_{s}}$ is always relevant and therefore increases
to strong coupling. Looking at Eq. \eqref{eq:kdependece}, we see
that this is sufficient to ensure that $K_{s}$ flows to zero, independent
of what happens to the charge tunneling. The charge tunneling term
will therefore always become relevant at some energy (temperature)
scale, which we will denote $T^{*}$ and discuss below. In this parameter
regime, the eventual endpoint of the RG flow is then the stable CC
phase.

\begin{figure}
\includegraphics[width=1\columnwidth]{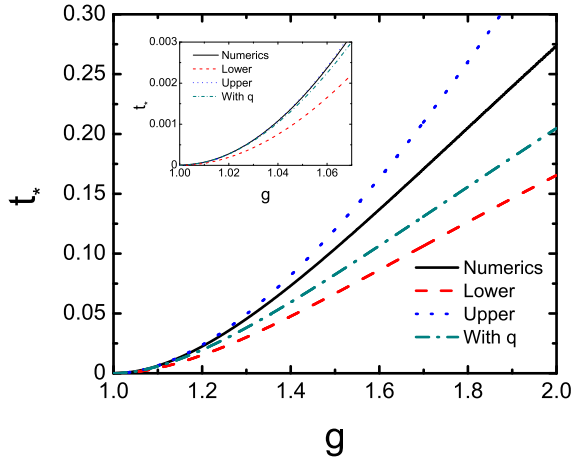} \caption{\label{fig:CC-and-II}Phase transition boundary between CC and II
phases in region $1<g<2$. The figure includes both the upper and
lower bounds in Eq. \eqref{eq:tcbound} obtained analytically, as
well as the results obtained by numerically integrating the RG equations
\eqref{eq:resonnatOurAll} both without ($A=0$) and with ($A=1$)
the quadratic terms. The inset zooms into the region around $g\rightarrow1$
where the critical $t_{*}\rightarrow0$.}
\end{figure}

For $1<g<2$, the situation is different, as the initial flow of both
$\tilde{t}_{_{c}}$ and $\tilde{t}_{_{s}}$ is towards zero. The flow
of $K_{s}$ depends on the magnitude of $\tilde{t}_{_{c}}$ and $\tilde{t}_{_{s}}$
-- meaning that if these are small (and become smaller under RG),
then $K_{s}$ might never decrease below the value needed to turn
one of the tunneling terms to be relevant. In other words, there is
a phase transition boundary separating an II phase from the CC one.
We note that in contrast to the case of a single constriction analyzed
in Ref.~\onlinecite{Teo2009} which found a similar phase boundary
at \textit{intermediate coupling} (also included in Fig.~\ref{fig:RG_double_ITI}
for reference), the present separatrix occurs at \textit{weak coupling}
and therefore can be fully analyzed in the context of the RG equations
in Eqs. \eqref{eq:resonnatOurAll}.

Setting $g=1+\varepsilon$ with $\varepsilon\ll1$, and for convenience
assuming the bare tunneling strengths are equal $\tilde{t}_{_{c}}=\tilde{t}_{_{s}}$
we analytically determine bounds on the critical $t_{*}$ separating
flow to the II phase from flow to the CC phase (see Appendix \ref{sec:Analytical}):
\begin{equation}
\frac{\varepsilon^{2}}{2}<t_{*}<\frac{\varepsilon^{2}}{\sqrt{2}}.\label{eq:tcbound}
\end{equation}
 In other words, as $g\rightarrow1$ from above, the critical tunneling
strength approaches zero quadratically. This is compared with the
true result obtained by numerical integration of Eq.~(\ref{eq:resonnatOurAll})
in Fig.~\ref{fig:CC-and-II}.

The above analysis was done ignoring the quadratic terms in Eq. \eqref{eq:resonnatOurAll},
i.e. setting $A=0$. However exactly along the separatrix, the linear
term of the equations becomes zero, and therefore it is not \textit{a
priori} obvious that one may ignore the quadratic terms. As it turns
out though (see Appendix \ref{sec:Analytical}), the quadratic nature
of $t_{*}$ as shown in Eq. \eqref{eq:tcbound} means that along the
separatrix, the quadratic terms still remain small, and therefore
do not strongly affect the position of the boundary line -- in other
words, it still goes to zero quadratically as $g\rightarrow1$. This
conclusion is backed up by again numerically solving the RG equations
with the quadratic terms $A=1$; this is also plotted in Fig.~\ref{fig:CC-and-II}.


We can now understand the full phase diagram of the model of a resonant
double constriction as shown in Fig.~\ref{fig:RG_double_ITI}. For
$1/2<g<1$, the system always flows to a CC phase, while for $1<g<2$
there is a phase boundary as a function of the strength of the constrictions
(controlled by an appropriate gate voltage) between the CC and II
phases. Going to stronger interactions and without presenting details,
we also find a stable IC phase for $g<1/2$ and a stable CI phase
for $g>2$. The calculations are exactly analogous to those done for
a single constriction in Ref.~\onlinecite{Teo2009}.

In the language of Kondo physics, the CC phase corresponds to the
one-channel Kondo fixed point, while IC is the two-channel Kondo fixed
point (see Appendix \ref{sec:KondoHamiltonian}). We therefore conclude
that the two-channel Kondo fixed point is stable only for $g<1/2$,
in strong contrast to recent reports\cite{Law2010,Lee2013} that the
elusive two-channel Kondo fixed point might be accessible for all
$g<1$. One possible explanation for this discrepancy is that there
is an important difference in the models of those works and that of
ours, to do with the size of $J_{1}^{z}$ {[}see Eq.~(\ref{eq:kondostandard})
in appendix A{]}. We ignore this term {[}see Eq. (\ref{eq:J1z}){]},
as it does not directly affect transport, and it is marginal so it
does not become large under RG flow. On the other hand, the work of
Ref.~\onlinecite{Law2010} analyzes the stability of Kondo phases
by finding exactly solvable (Toulouse) points, which require a large
$J_{1}^{z}$. In fact, it has already been advocated by Chung and
Silotri\cite{Chung2012} that there is a quantum phase transition
between the one- and two-channel Kondo fixed points in this parameter
region. Our work supports this scenario.

\subsection{Temperature dependence of conductance for resonant double impurity}

\begin{figure}
\begin{centering}
\includegraphics[width=1\columnwidth]{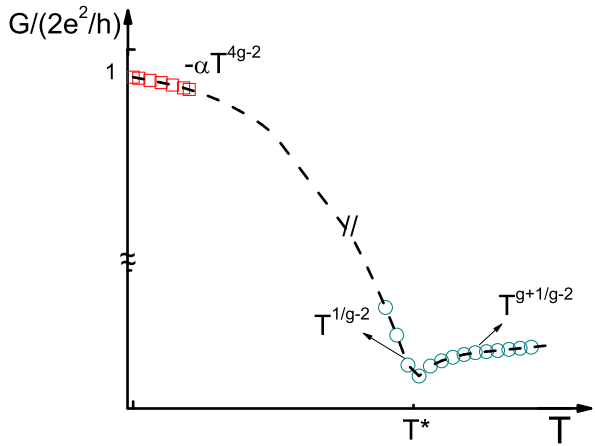} 
\par\end{centering}

\caption{\label{fig:Linear-conductance-as}Conductance as a function of temperature
at $1/2<g<1$ ($g=0.75$ in this plot). In the high temperature regime,
there is a power law $G\sim T^{g+1/g-2}$, at intermediate temperature,
$G\sim T^{1/g-2}$, and close to zero temperature, $\delta G\sim-T^{4g-2}$
where $\delta G$ is the deviation from perfect conductance. Data
points close to the bottom of the graph are not scaled to the data
points close to the full conductance.}
\end{figure}

\begin{figure}
\includegraphics[width=1\columnwidth]{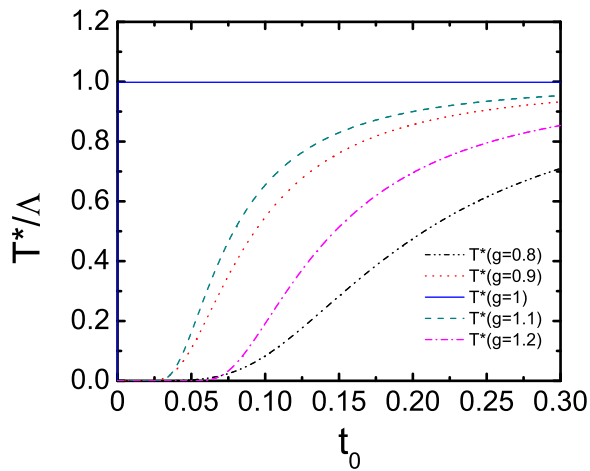} \caption{\label{fig:Kondo-temperature-as} The temperature of minimum conductance
$T^{*}$ as a function of bare tunneling strength $t_{0}\equiv\tilde{t}_{c}(l=0)$
for various interaction parameters g. }
\end{figure}

For most of the phase diagram with $1/2<g<2$ the fixed point of the
flow is the conducting one. This means that at $T=0$, one finds perfect
conductance through the resonant double impurity. However, if the
constrictions are strong, or in other words the conductance at high
temperatures is small, then the charge conductance as a function of
temperature is non-monotonic. To see this, we recall that the conductance
will be proportional to the square of the (renormalized) tunneling
strength at the appropriate energy scale 
\begin{equation}
G(T)\propto\frac{\tilde{t}_{_{c}}^{2}(l)}{1+\tilde{t}_{_{c}}^{2}(l)},
\end{equation}
 where $l=\ln\Lambda/T$. In principle, there is also a term proportional
to the electron hopping term $t_{_{e}}^{2}$, however as this is always
irrelevant it doesn't play an important role in the following discussion.

Now, while the conducting phase means $\tilde{t}_{_{c}}\rightarrow\infty$
as $l\rightarrow\infty$ (i.e. $T\rightarrow0$), the initial flow
of this parameter is in the other direction. By numerically integrating
the flow equation \eqref{eq:resonnatOurAll} one can obtain $G(T)$,
as shown in Fig.~\ref{fig:Linear-conductance-as} for a representative
value of parameters. In the high temperature regime, the temperature
dependence has a form $G\sim T^{g+1/g-2}$, while close to zero temperature,
the conductance is determined by the stable CC phase where the correction
to the quantized conductance is given by $\delta G\sim-T^{4g-2}$.
The most interesting feature of this plot however is the minimum at
some flow scale $l^{*}=\ln\Lambda/T^{*}$. This is the scale at which
the $\tilde{t}_{_{c}}$ operator changes from being irrelevant to
being relevant. In physical terms, this temperature give the scale
at which the crossover between two channel and one channel Kondo physics
(as discussed in the previous subsection) occurs. At temperatures
(or energy scales) higher than $T^{*}$, the physics appears to be
that of the two channel Kondo model --- however at lower temperatures,
the one channel Kondo physics takes over.

Numerically, $T^{*}$ can be determined as a function of the bare
tunneling $t_{0}\equiv\tilde{t}_{_{c}}$ and Luttinger constant $g$,
as shown in Fig. \ref{fig:Kondo-temperature-as}. We can also obtain
analytic expressions for $T^{*}$ in various limits (see Appendix
\ref{sec:RGflow}). If $g=1+\varepsilon$, we find in the limit $\varepsilon/t_{0}\ll1$:
\begin{equation}
T^{*}\sim\Lambda\left(1-\frac{\varepsilon^{2}}{2t_{0}^{2}}\right).\label{eq:Tstar1}
\end{equation}
 For any real material, the cutoff $\Lambda$ may be taken to be the
bulk energy gap. For HgCdTe this is about 40 meV, meaning that the
temperature of minimum conductance is of the order of one hundred
Kelvin, so long as the bare (high temperature) conductance is not
too small.

In the opposite limit of $\varepsilon/t_{0}\gg1$, and $g<1$, we
find 
\begin{equation}
T^{*}\sim\Lambda t_{0}^{\frac{1}{1-g}},\label{eq:Tstar2}
\end{equation}
 with the minimum conductance $G_{\mathrm{min}}=G(T^{*})$ given by
\begin{equation}
G_{\mathrm{min}}\sim t_{0}^{\frac{g+1}{g}}.\label{eq:Gmin}
\end{equation}
 Such a limit does not exist for $g>1$, as there is no $T^{*}$ within
the II phase. However, we can say that close to the phase transition
on the CC side, $T^{*}$ is very small, going to zero exactly at the
phase boundary.

At temperature below $T^{*}$, when the system now behaves like the
one channel Kondo model, there is another important temperature --
the Kondo temperature $T_{K}$. This is defined as the energy scale
when $\tilde{t}_{_{c}}(T_{K})=1$; or in other words, the strong coupling
regime is reached. As we will show in the next subsection, this energy
scale is crucial when describing the physics of the double constriction
tuned slightly off resonance. We will therefore now briefly discuss
the Kondo temperature in our model, for a derivation of these results
see Appendix \ref{sec:RGflow}. When $g$ is close to or greater than
$1$, there is not a great separation of energy scales between $T^{*}$
and $T_{K}$. Hence to within prefactors of the order of unity, these
energy scales are the same. However, under the same conditions of
validity as Eq.~\eqref{eq:Tstar2}, we find that 
\begin{equation}
T_{K}\sim T^{*}\, t_{0}^{\frac{1+g}{2g-1}}=\Lambda t_{0}^{\frac{g(2-g)}{(2g-1)(1-g)}},\label{eq:Tkondo}
\end{equation}
 and therefore there is a parametrically large region between the
minimum of conductance, $T^{*}$, and the strong coupling regime $T_{K}$,
where the physics of the CC point takes over. In this intermediate
temperature range, there is a third power law (beyond the high and
low temperature limits mentioned previously), $G\sim T^{1/g-2}$.
In a system with sufficiently strong constrictions, all three of these
power laws should be seen clearly, giving an experimental signature
for the presence of the helical LL edge states, as well as a consistency
check for the experimental determination of $g$. The main downside
of this scheme is that the stronger the constriction (and therefore
the wider the power law regions), the lower the temperature scales
relevant for the crossovers.

\subsection{Double constriction tuned slightly off resonance}

\label{sec:off-resonance}

\begin{figure}
\includegraphics[width=0.8\columnwidth]{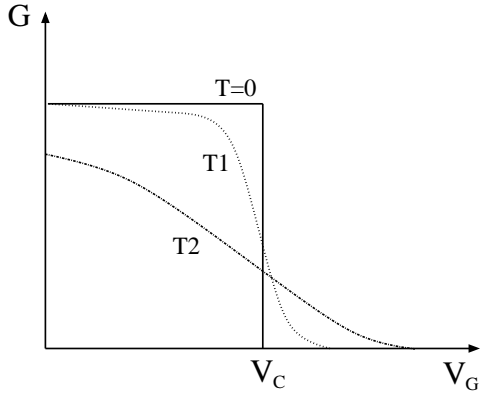} \caption{\label{fig:MITMetal-insulator-transistion} Schematic plot of the
metal insulator transition: conductance as a function of detuning
from resonance $V_{G}$. At zero temperature, there exhibits a sharp
metal insulator transition at $V_{C}$, but at finite temperatures
this sharp transition becomes a crossover.}
\end{figure}


We now add a twist to the problem of a double constrictions by considering
that the resonant condition in Eq.~\eqref{eq:resonance_condition}
is almost, but not quite, satisfied. This may be tuned in an experiment
by yet another gate capacitively coupled to the dot, and therefore
changing the equilibrium charge in the dot away from the odd integer
required for resonance. We introduce a physical parameter $V_{G}$,
which can be controlled by a top gate over the dot, to quantify the
distance from resonance, in other words $V_{G}$ is the energy difference
between the two lowest energy states within the dot.

If $V_{G}$ is very large (i.e. the system is far from resonance),
then there are no internal dynamical processes within the dot, and
the double constriction looks identical to a single constriction with
some effective tunneling across it. This motivates the use of a two-cutoff
RG procedure. At energy scales larger than $V_{G}$, the system doesn't
sense the perturbation and it looks exactly like the case of resonance.
Therefore the RG equations in Eq. \eqref{eq:resonnatOurAll} for the
resonant case are used. However when the energy scale is lower than
$V_{G}$ , the system behaves like a single constriction, so the RG
equations are switched to those for a single constriction {[}see Eqs.~(\ref{eq:generalRG})
and (\ref{eq:scaling_dimensions_t}){]}.

For $1/2<g<2$, we show in Fig. \ref{fig:RG_double_ITI}, the phase
boundary lines between insulating and conducting phases for both the
resonant double constriction case (as discussed above), and the single
constriction case (after Ref.~\onlinecite{Teo2009}). By far the
most interesting region of the phase diagram is the large region of
phase space between the red dashed and green solid lines where the
resonant constriction conducts, while the single constriction is an
insulator. In other words, if the detuning from resonance is introduced
then the system is a conductor at $V_{G}=0$ but an insulator when
$V_{G}$ is large. This implies that there must be a critical detuning
$V_{C}$ where the system undergoes a metal-insulator transition.
Of course, this is a transition only at zero temperature $T=0$, at
non-zero temperatures it becomes a crossover as shown schematically
in Fig.~\ref{fig:MITMetal-insulator-transistion}.

\begin{figure}
\includegraphics[width=1\columnwidth]{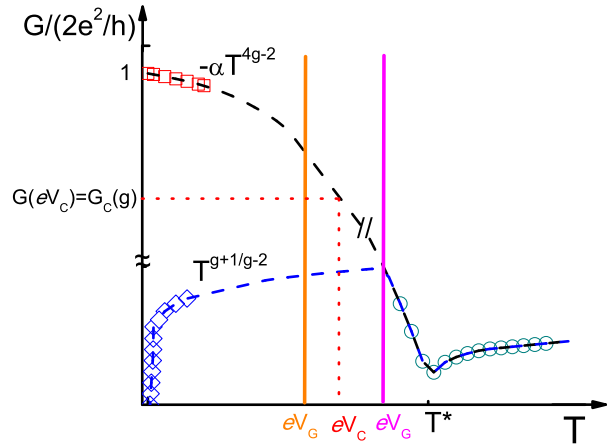}\caption{\label{fig:CoffConductance-vs-temperature}Conductance vs temperature
at small detuning from resonance, $V_{G}$. For $T\gg V_{G}$, the
system is described as if it were resonant, but for $T\ll V_{G}$
the system effectively looks like a single constriction. When $V_{G}$
is smaller than some critical $V_{C}$ the system is conducting otherwise
the system flow is insulating. Numerical data obtained by integrating
the RG equations is represented by blue squares, red diamonds, and
green circles (Data points close to the bottom of the graph are not
scaled to the data points close to perfect conductance).}
\end{figure}

The value of $V_{C}$ may be estimated via the two-cutoff RG procedure
outlined above. Basically, the bare parameters flow initially under
the resonant RG equations in Eq. \eqref{eq:resonnatOurAll} from an
energy scale (or temperature) $\Lambda$ down to $V_{G}$. At this
energy scale, one switches to the physics of the single constriction
-- so in order for the system to remain metallic, the renormalized
conductance at this energy scale must be bigger than the critical
conductance $G_{C}(g)$ for the metal-insulator transition in the
single constriction case, as found by Teo and Kane in Ref.~\onlinecite{Teo2009}.
Hence the critical $V_{C}$ is given by the matching condition 
\begin{equation}
G(T=eV_{C})=G_{C}(g).
\end{equation}
 When $V_{G}<V_{C}$, the system flows to conducting fixed point,
while for $V_{G}>V_{C}$, the system flows to insulating fixed point.

As can be seen from Fig.~\ref{fig:RG_double_ITI}, the critical conductance
for the single constriction $G_{C}(g)\sim1/2$, so long as $g$ is
not too close to either $1/2$ or $2$; in other words, the phase
boundary lies at intermediate coupling. Now, the energy scale at which
this happens is exactly the Kondo temperature, as we defined it in
the previous section. Hence, over a wide range of values of interaction
$g$ and up to numerical prefactors of the order of unity, $V_{C}\sim T_{K}/e$.
Hence by detuning the constrictions from the resonance transition,
one makes a direct probe of the Kondo temperature of the system.

The behavior of conductance as a function of temperature is shown
for both these cases in Fig.~\ref{fig:CoffConductance-vs-temperature}.
It is worth commenting that for $V_{G}>V_{C}$, the conductance as
a function of temperature shows a local maximum at $T\sim eV_{G}$.
This is on top of the local minimum that occurs at $T=T^{*}$ due
to Kondo physics, giving a rather interesting profile to the $G-T$
characteristics of the system, which should be observable experimentally.


\begin{figure}
\includegraphics[width=1\columnwidth]{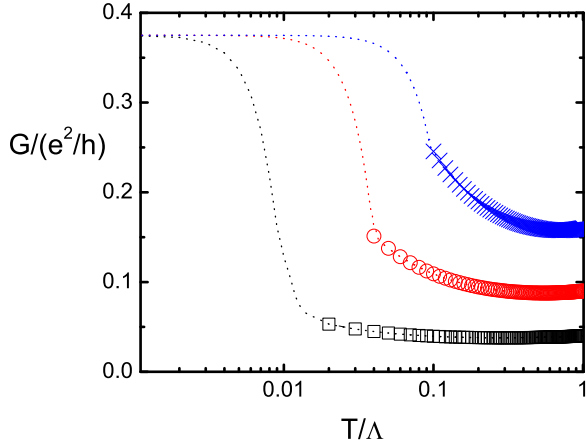} \caption{Conductance on the critical line $V_{G}=V_{C}$. When the system is
tuned to this condition, the single-constriction metal-insulator phase
boundary acts as an attractive fixed fixed. Here we plot three various
initial tunneling strength along this critical line. The ultimate
conductance at $T=0$ is given by $G_{C}(g)$. In this plot, $g=0.75$.}

\label{fig:criticalline} 
\end{figure}

Finally, we mention the interesting possibility of tuning the system
exactly to the critical line where $V_{G}=V_{C}$. With this condition
satisfied, the metal-insulator boundary found in Ref.~\onlinecite{Teo2009}
acts as an \textit{attractive} fixed point; the ultimate $T=0$ conductance
is given by this value. Some typical plots of conductance as a function
of temperature on this critical line are given in Fig.~\ref{fig:criticalline}.
We leave a full analysis of the scaling behavior expected here for
future work.


\section{\label{sec:2d-fractional-TIs} 2d fractional topological insulators}

We now turn to the case of adding constrictions to 2D FTIs with a
filling factor $\nu=1/m$ ($m$ is an odd integer). We first discuss
the appropriate model of such a state of matter, before adding a single
constriction and then a resonant double constriction as discussed
above for the integer case. The analysis will be very similar to the
integer case: first bosonize the model in the absence of constrictions,
then add constrictions and use an RG analysis to study the stability
of the weak and strong coupling fixed points, which allows us to draw
the phase diagram. We will therefore concentrate on what changes when
one moves from the integer to the fractional cases; for details of
the calculations, see the previous sections.


As proposed by Levin and Stern,\cite{Levin2012} a toy model of a
2D FTI is made from two copies of a fractional quantum Hall state
with opposite spin species, and a short range interaction between
them. Within this model, the filling factor $\nu$ modifies the Hamiltonian
in Eq. (\ref{eq:topedge}) as follows:\cite{Wen1990,Wen1992} 
\begin{eqnarray}
H^{T(B)} & = & \int\mathtt{d}x\left[\frac{\pi v_{F}}{\nu}\left(\rho_{R\uparrow(\downarrow)}^{2}+\rho_{L\downarrow(\uparrow)}^{2}\right)\right.\label{eq:fratopedge}\\
 & + & \left.\lambda_{2}\rho_{R\uparrow(\downarrow)}\rho_{L\downarrow(\uparrow)}+\lambda_{4}\left(\rho_{R\uparrow(\downarrow)}^{2}+\rho_{L\downarrow(\uparrow)}^{2}\right)\right],\nonumber 
\end{eqnarray}
 where the density $\rho$ is now given by 
\begin{equation}
\rho=\frac{\sqrt{\nu}}{2\pi}\partial_{x}\Phi_{\eta\sigma},
\end{equation}
 in terms of boson fields $\Phi_{\eta\sigma}$ satisfying the commutation
relation in Eq.~\eqref{eq:kacmoody}. The electron creation operator
is modified to 
\begin{equation}
\Psi_{e,\eta\sigma}^{\dagger}=e^{-i\frac{\Phi_{\eta\sigma}}{\sqrt{\nu}}},\label{eq:electronoperator-1}
\end{equation}
 and the quasiparticle creation operator with fractional charge $\nu e$
is 
\begin{equation}
\Psi_{q,\eta\sigma}^{\dagger}=e^{i\sqrt{\nu}\Phi_{\eta\sigma}}.\label{eq:quasioperator-1}
\end{equation}
The chiral boson fields can further be expressed as before by Eq.
\eqref{eq:veryfield}.

Going through these transformations, the Hamiltonian in Eq.~\eqref{eq:fratopedge}
once more splits into a spin and charge part as written in Eq.~(\ref{eq:fourth})
but with a modification of the parameters: 
\begin{equation}
g=\sqrt{\frac{1+\lambda_{4}/\pi\nu v_{F}-\lambda_{2}/2\pi\nu v_{F}}{1+\lambda_{4}/\pi\nu v_{F}+\lambda_{2}/2\pi\nu v_{F}}},
\end{equation}
 and 
\begin{equation}
v=v_{F}\sqrt{\left(1+\frac{\lambda_{4}}{\pi\nu v_{F}}\right)^{2}-\left(\frac{\lambda_{2}}{2\pi\nu v_{F}}\right)^{2}}.
\end{equation}
 Note that if we put $\nu=1$, we recover the expressions in Eqs.~\eqref{eq:velocity}
and \eqref{eq:g} of the integer case.

\subsection{Single constriction}

We now add a single constriction (with geometry as shown in Fig.~\ref{fig:openup})
to the FTI. Before plugging the modified operators in Eqs.~\eqref{eq:electronoperator-1}
and \eqref{eq:quasioperator-1} into the previous expressions, there
is one final point to mention, which is under which circumstances
may fractionally charged quasi-particles be backscattered (or tunnel),
and in which cases are only electrons allowed. This has recently been
discussed by Beri and Cooper in Ref.~\onlinecite{Beri2012} for
magnetic impurities in a single edge, where the two species (spin
up and spin down) of electrons or quasi-particles must scatter into
each other. The present case is much simpler, as the scattering between
the top and lower edges preserves the species index of the particles
being scattered. Consequently, we can use the conventional wisdom
from fractional quantum Hall systems\cite{Sarma1997} and observe
that the scattering from the upper to lower edges in the weak backscattering
limit may be quasiparticles, while only electrons may tunnel between
the left and right of a system that has been pinched off.

With this information, we are now ready to modify the calculations
presented above for the fractional case. In the weak backscattering
limit, quasiparticles are scattered between the upper and lower edges.
Substituting the quasi-particle operator in Eq. (\ref{eq:quasioperator-1})
for $\Psi$ in Eq.~(\ref{eq:backscatter}), we obtain the quasiparticle
backscattering Hamiltonian 
\begin{multline}
H_{b}^{F}=\frac{\upsilon_{_{e}}^{q}}{2\pi\alpha}\cos\sqrt{\nu}\theta_{c}\cos\sqrt{\nu}\theta_{s}+\frac{\upsilon_{_{c}}^{q}}{2\pi\alpha}\cos2\sqrt{\nu}\theta_{c}\\
+\frac{\upsilon_{_{s}}^{q}}{2\pi\alpha}\cos2\sqrt{\nu}\theta_{s},
\end{multline}
 where $\upsilon_{_{e}}^{q}$ stands for the \textit{single quasiparticle}
backscattering process across the QPC, $\upsilon_{_{c}}^{q}$ represents
a quasiparticle pair backscattering with opposite spins, and $\upsilon_{_{s}}^{q}$
represents the transfer of 2$\nu e$ charged particles from the top
to the bottom edges. The leading order renormalization group flows
for each process is then 
\begin{equation}
\frac{d\upsilon_{_{a}}^{q}}{dl}=\left(1-\Delta_{\upsilon a}^{(q)}\right)\upsilon_{_{a}}^{q},
\end{equation}
 with scaling dimensions \begin{subequations} 
\begin{eqnarray}
\Delta_{\upsilon e}^{(q)} & = & \frac{\nu}{2}\left(g+g^{-1}\right),\\
\Delta_{\upsilon c}^{(q)} & = & 2\nu g,\\
\Delta_{\upsilon s}^{(q)} & = & 2\nu g^{-1}.
\end{eqnarray}
 \end{subequations}

On the contrary, in the weak link limit, quasiparticle tunneling between
the two fractional fluids is forbidden, so the tunneling Hamiltonian
involves only electron operators 
\begin{multline}
H_{t}^{F}=\frac{t_{_{e}}}{2\pi\alpha}\cos\frac{\bar{\phi}_{c}}{\sqrt{\nu}}\cos\frac{\bar{\phi}_{s}}{\sqrt{\nu}}+\frac{t_{_{c}}}{2\pi\alpha}\cos\frac{2\bar{\phi}_{c}}{\sqrt{\nu}}\\
+\frac{t_{_{s}}}{2\pi\alpha}\cos\frac{2\bar{\phi}_{s}}{\sqrt{\nu}}.
\end{multline}
 Here the labeling is the same as the integer case. The scaling dimensions
of these operators are: \begin{subequations} 
\begin{eqnarray}
\Delta_{te}^{(q)} & = & \frac{1}{2}\nu^{-1}\left(g+g^{-1}\right),\\
\Delta_{tc}^{(q)} & = & 2\nu^{-1}g^{-1},\\
\Delta_{ts}^{(q)} & = & 2\nu^{-1}g.
\end{eqnarray}
 \end{subequations}

\begin{figure}
\includegraphics[width=1\columnwidth]{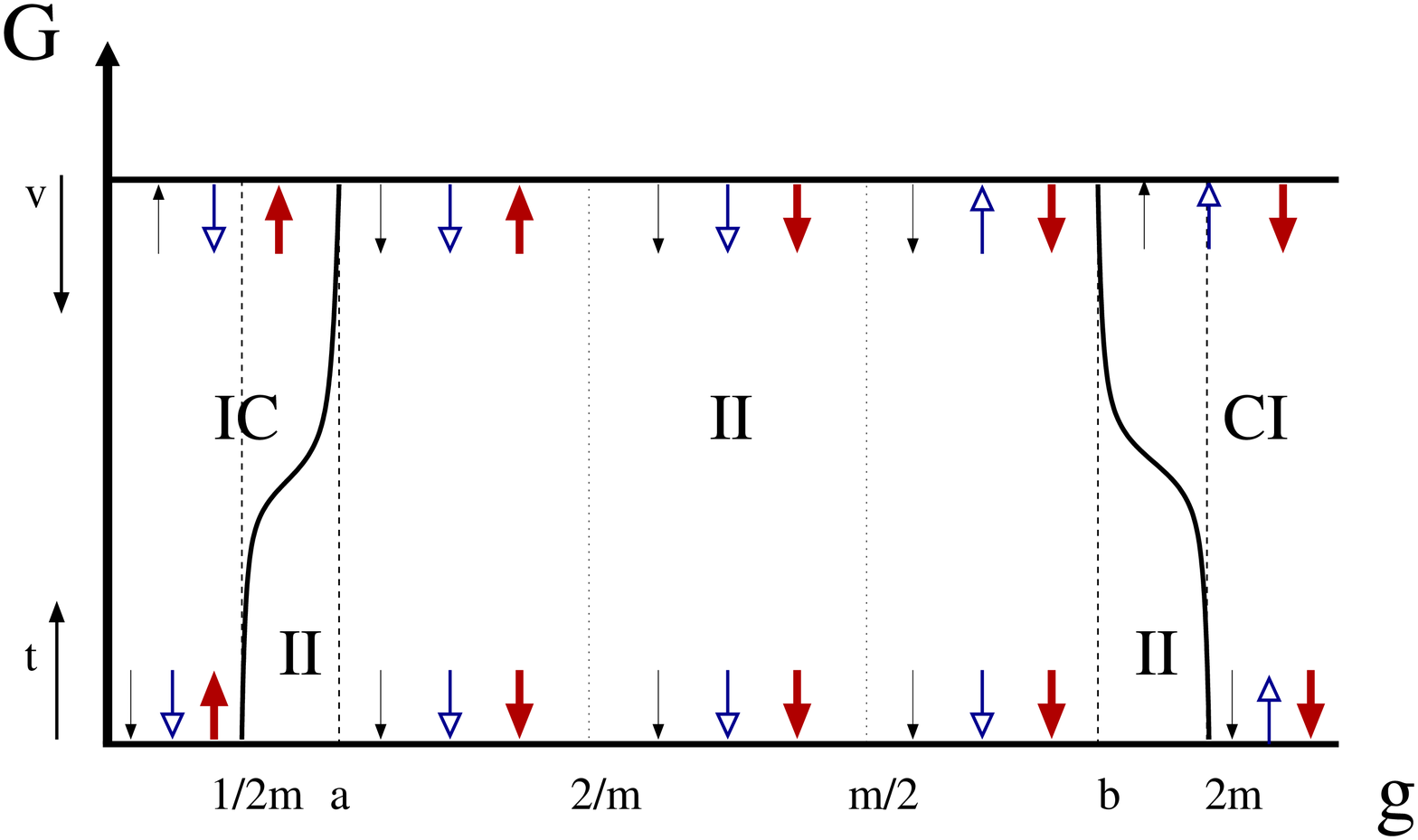} \caption{\label{fig:cc}Phase diagram for a point contact in a FTI as a function
of Luttinger constant $g$. The points $a=m-\sqrt{m^{2}-1}$ and $b=m+\sqrt{m^{2}-1}$.
The result for the weak backscattering limit is shown on the upper
part of the figure, and the result for the weak link limit is shown
in the lower part of the figure. Thin black arrows represent single
electron processes, blue arrows (with hollow arrow heads) represent\textit{
charge} tunneling (or backscattering) and thick red arrows represent
\textit{spin} tunneling (or backscattering) events. The black solid
curves indicate schematically the phase boundary between the IC and
II phases for $1/2m<g<a$, and between the CI and II phases for $b<g<2m$.
A stable insulating phase is found for all $a<g<b.$ }
\end{figure}


The phase diagram for a 2D FTI is obtained by analyzing the stability
of these two limits, and is plotted in Fig. \ref{fig:cc}. As shown
in the upper panel of the figure (weak backscattering limit), the
single particle backscattering operator with strength $\upsilon_{e}^{q}$
becomes relevant when $m-\sqrt{m^{2}-1}<g<m+\sqrt{m^{2}-1}$, which
doesn't happen in the case of integer QSHIs. The operator coupled
to $\upsilon_{c}^{q}$ is relevant when $g<m/2$ , and that coupled
to $\upsilon_{s}^{q}$ becomes relevant when $g>2/m$. In the other
limit on the bottom panel of the plot, the parameter $t_{e}$ is irrelevant
for any $g$, while $t_{c}$ becomes relevant when $g>2m$ and $t_{s}$
is relevant for $g<1/2m$. We find that in the most likely physical
regime of $g$ not too far from one, there exists a stable insulating
phase. It coincides with the prediction in the fractional quantum
Hall effect in which a even a weak impurity will drive the fractional
quantum Hall fluid to an insulating phase.\cite{Sarma1997}

\subsection{\label{sec:2d-fractional-TI-2}Resonant double constriction}

Finally, we consider a resonant double constriction for a FTI, with
a geometry as shown in Fig.~\ref{fig:doubleQPC}. As before, in the
weak backscattering limit one considers quasiparticle processes, and
the effective action in Eq.~(\ref{eq:dualmap}) is modified by $\theta\rightarrow\sqrt{\nu}\theta$.
Again, the resonant condition means that the single particle backscattering
process is absent, and the scaling dimensions for the two particle
processes are as for the single constriction case \begin{subequations}
\begin{eqnarray}
\Delta_{\upsilon c}^{(q)} & = & 2\nu g,\\
\Delta_{\upsilon s}^{(q)} & = & 2\nu g^{-1}.
\end{eqnarray}
 \end{subequations}

For the weak link limit, only electron processes are allowed, which
means $\bar{\phi}$ is modified by $\bar{\phi}/\sqrt{\nu}$ in Eq.
(\ref{eq:kondotunneling}). The scaling dimensions for the three tunneling
processes are as follows: \begin{subequations} 
\begin{eqnarray}
\Delta_{te}^{(q)} & = & \frac{1}{2\nu}\left(g+g^{-1}\right),\\
\Delta_{tc}^{(q)} & = & \frac{1}{2\nu}\left(K_{s}g+g^{-1}\right),\\
\Delta_{ts}^{(q)} & = & \frac{g}{2\nu}\left(1+K_{s}\right).
\end{eqnarray}
 \end{subequations} The way that the full RG equations (including
the flow of $K_{s}$) given for the integer case in Eq. \eqref{eq:resonnatOurAll}
is modified is clear from the above expressions. We also note that
these \textit{electronic} tunneling processes are still correctly
given by the instantons of the potential in Eq. \eqref{eq:kondotunneling}
with the \textit{quasiparticle} modification above; meaning that our
picture of quasiparticles and electrons is a consistent one.

\begin{figure}
\includegraphics[width=1\columnwidth]{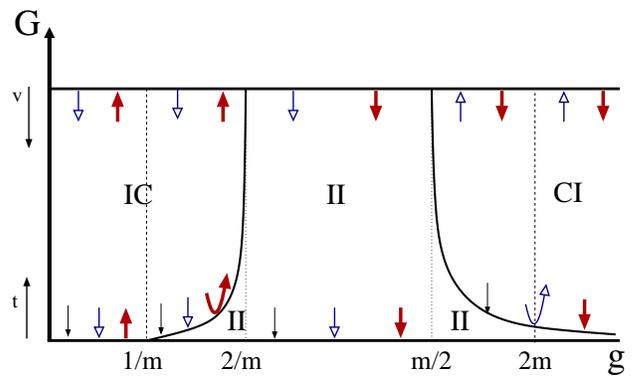} \caption{\label{fig:RG-double-FTI}RG flow for a double barrier in a FQSHI.
The result for the weak backscattering limit is shown on the upper
part of the figure: blue (with hollow arrow heads) and red arrows
are flows for processes involving two charges and two spins respectively.
The result for the weak link limit is shown in the lower part of the
figure, where the thin black arrows represent single electron tunneling
events, blue arrows (with hollow arrow heads) represent \textit{single
charge} tunneling and thick red arrows represent \textit{single spin}
tunneling events. The shapes of the thick black phase boundary lines
between the IC, II and CI phases are schematic. In the regions with
curved arrows, we predict non-monotonic temperature dependence of
conductance: charge conductance in the CI phase, and spin conductance
in the IC phase (with g>1/m). 
}
\end{figure}

The RG flow is summarized in Fig.~\ref{fig:RG-double-FTI}, where
we predict an insulating phase occurs when $2/m<g<m/2$. At $g<2/m$
there is a transition to the IC phase, while a transition to the CI
phase occurs at $g>m/2$ -- although neither of these transition lines
is vertical. In the CI phase, we predict a non-monotonic temperature
dependence of charge conductance, while the IC phase exhibits non-monotonic
temperature dependence of spin conductance, so long as $g>1/m$. We
note that unlike the integer case, all this interesting behavior happens
at strong interaction strengths, of $g$ not close to one. 

\section{\label{sec:Conculding-remarks}Summary}

In summary, we have considered a quantum spin Hall insulator with
a double constriction as shown in Fig.~\ref{fig:doubleQPC}. The
transport properties of the setup are controlled by three parameters:
the interaction strength within the material parametrized as the Luttinger
liquid constant $g$, strength of the constrictions $\upsilon$, and
how close the system is to resonance $V_{G}$. The former is an intrinsic
property of the material (although may be partially controlled via
screening through metal gates in close proximity), however the latter
two may be easily controlled experimentally through appropriate side
and top gates on the sample.

We find that if $1/2<g<1$ the system is always conducting at resonance,
though with conductance non-monotonic as a function of temperature
due to Kondo physics. Unlike some recent reports however, we do not
find the two-channel Kondo fixed point stable in this region -- this
may be due to an interaction between the dot and the leads which we
consider to be small, while previous studies have taken it to be large.
We believe the most likely scenario is a quantum phase transition
as a function of this interaction, however more work needs to be done
in this direction.

Our conducting phase may be driven to an insulating phase by tuning
$V_{G}$, with a metal-insulator transition at some critical $V_{G}=V_{C}$.
On the insulating side of the transition, the conductance as a function
of temperature has both a maximum and a minimum (see Fig.~\ref{fig:CoffConductance-vs-temperature});
we believe this to be a fascinating experimental signature of the
physics discussed in this work.

Finally, we studied the same geometry for the as yet hypothetical
fractional topological insulators with filling fraction $\nu=1/m$,
$m$ being an odd integer. In this case, we find that all of the exciting
physics of the integer case has moved to much higher interaction strengths;
for $2/m<g<m/2$ we predict a stable insulating phase. It appears
to be a curious characteristic of the fractional TI edge states that
while single edges show a remarkable resilience to single edge perturbations,\cite{Beri2012}
they are completely unstable to a coupling between two edges.


\begin{acknowledgments}
We thank R. Berkovits, B. Beri, N. Cooper, W. Metzner and I. Safi
for useful discussions. This work was supported by the Israeli Science
Foundation under Grants No. 819/10 (D.G) and No. 599/10 (E.S), the
US-Israel Binational Science Foundation (BSF) under Grant No. 2008256,
the German-Israeli Foundation under Grant No. 1167/2011, and SPP 1285
and SPP 1666 of the Deutsche Forschungsgemeinschaft. 
\end{acknowledgments}

\appendix

\section{\label{sec:KondoHamiltonian}Derivation of the Kondo Hamiltonian}

As in the case of a single constriction, if the constrictions are
very large then it is more convenient to begin from the completely
pinched off limit and add in the weak tunneling between the leads
and the dot perturbative. The tunneling Hamiltonian takes the form
\begin{equation}
H_{tun}=T_{L}\sum_{\sigma}\Psi_{1\sigma}^{\dagger}\Psi_{d\sigma}+T_{R}\sum_{\sigma}\Psi_{2\sigma}^{\dagger}\Psi_{d\sigma}+h.c.,\label{eq:tunnelinghamil}
\end{equation}
 where $\Psi_{1\sigma}$, $\Psi_{2\sigma}$, and $\Psi_{d\sigma}$
annihilate an electron with spin $\sigma$ on the left, right leads
and the dot respectively, and we have allowed different tunneling
amplitudes $T_{L}$ and $T_{R}$ through the left and right constriction
respectively. However this is not the full story, as the resonance
condition in Eq. \eqref{eq:resonance_condition} ensures that the
dot has a \textit{fixed odd number} of electrons sitting on it, with
some charging energy $U$ to add (or remove) an electron from it.
The odd number means that there is still an internal degree of freedom
on the dot -- the spin -- which has two degenerate states {[}see Eqs.~(\ref{eq:veffkondo})
and (\ref{eq:resonance_condition}){]}.

Restricting ourselves to low energies, one therefore looks at the
second order processes in which the final state of the dot remains
within the low energy manifold. Applying second order perturbation
theory generates the following four operators: \begin{subequations}\label{eq:secondorder}
\begin{eqnarray}
H_{e}^{K} & = & -\frac{T_{L}T_{R}^{*}}{U}\sum_{\sigma}\Psi_{1\sigma}^{\dagger}\Psi_{2\sigma}\left(1-\Psi_{d\sigma}^{\dagger}\Psi_{d\sigma}\right)\\
 &  & \quad\quad-\frac{T_{L}^{*}T_{R}}{U}\sum_{\sigma}\Psi_{2\sigma}^{\dagger}\Psi_{1\sigma}\left(1-\Psi_{d\sigma}^{\dagger}\Psi_{d\sigma}\right),\nonumber \\
H_{c}^{K} & = & \frac{T_{L}T_{R}^{*}}{U}\sum_{\sigma}\Psi_{1\sigma}^{\dagger}\Psi_{2-\sigma}\Psi_{d-\sigma}^{\dagger}\Psi_{d\sigma}\\
 &  & \quad\quad+\frac{T_{L}^{*}T_{R}}{U}\sum_{\sigma}\Psi_{2\sigma}^{\dagger}\Psi_{1-\sigma}\Psi_{d-\sigma}^{\dagger}\Psi_{d\sigma},\nonumber \\
H_{s}^{K} & = & \frac{\left|T_{L}\right|^{2}}{U}\sum_{\sigma}\Psi_{1\sigma}^{\dagger}\Psi_{1-\sigma}\Psi_{d-\sigma}^{\dagger}\Psi_{d\sigma}\\
 &  & \quad\quad+\frac{\left|T_{R}\right|^{2}}{U}\sum_{\sigma}\Psi_{2\sigma}^{\dagger}\Psi_{2-\sigma}\Psi_{d-\sigma}^{\dagger}\Psi_{d\sigma}.\nonumber \\
H_{dd} & = & \frac{\left|T_{L}\right|^{2}}{U}\sum_{\sigma}\Psi_{1\sigma}^{\dagger}\Psi_{1\sigma}\Psi_{d\sigma}^{\dagger}\Psi_{d\sigma}\label{eq:J1z}\\
 &  & \quad\quad+\frac{\left|T_{R}\right|^{2}}{U}\sum_{\sigma}\Psi_{2\sigma}^{\dagger}\Psi_{2\sigma}\Psi_{d\sigma}^{\dagger}\Psi_{d\sigma}.\nonumber 
\end{eqnarray}
 \end{subequations} The physical meaning of each of the first three
terms is explained in the main text and schematically represented
in Fig.~\ref{fig:Resonant}. For the symmetrical case when both constrictions
are identical, $T_{L}=T_{R}=T_{0}$, and the bare $t_{\alpha}$ all
have transmission strengths proportional to $\left|T_{0}\right|^{2}$.
The fourth term, Eq.~\eqref{eq:J1z}, 
 takes the form of an $S^{z}S^{z}$ interaction. This term does not
transfer any spin or charge across the dot, and furthermore is marginal
under RG which we use as justification for neglecting it. However,
if for whatever reason the bare value of this fourth term is large,
its presence may have an important influence on the RG flow of the
other terms (see discussion in main text). A careful study of the
effect of this term is however beyond the scope of the present work.

Following the same procedure as for the single constriction case,
we then bosonize the first three tunneling terms above, arriving at
the answer 
\begin{multline}
H_{t}^{K}=\frac{t_{_{e}}}{2\pi\alpha}\cos\bar{\phi}_{+c}\cos\bar{\phi}_{+s}+\frac{\tilde{t}_{_{c}}}{2\pi\alpha}\cos\bar{\phi}_{+c}\cos\bar{\phi}_{-s}\\
+\frac{\tilde{t}_{_{s}}}{2\pi\alpha}\cos\bar{\phi}_{+s}\cos\bar{\phi}_{-s}.\label{eq:kondotunnelingappendix}
\end{multline}
 which is quoted in Eq.~(\ref{eq:kondotunneling}) in the main text.
Care must be taken however to understand the difference between the
fields $\bar{\phi}_{+c},\bar{\phi}_{+s}$ and the field $\bar{\phi}_{-s}$.
The first two are associated with tunneling of charge or spin from
the left to the right leads, and are no different from the equivalent
operators found in the single constriction case. The remaining field
however $\bar{\phi}_{-s}$ is associated with the state of the dot.
In fact, bosonizing the dot which is not a bulk system (in contrast
to the semi-infinite leads) is somewhat of a cheat; however continuing
this line of reasoning we are led to the following important point.
While a local operator in a one-dimensional wire can not renormalize
the Luttinger liquid constant of the wire, this is not true of the
operator $\cos\bar{\phi}_{-s}$ on the dot. Consequently, one should
define an \textit{effective Luttinger liquid parameter} of the dot,
$K_{s}g$ (where the bare value of $K_{s}=1$), which may change during
the renormalization procedure. This is one way to understand the parameter
$K_{s}$ appearing in the RG equations \eqref{eq:resonnatOurAll}
in the main text.

While the above argument for the introduction of $K_{s}$ turns out
to be mathematically correct, it clearly lacks rigor. This may be
rectified by looking at the problem from a different angle. By going
back to the original action in terms of the constrictions Eq.~(\ref{eq:dualmap})
and performing saddle point approximation on the second term,\cite{Rajaraman1982,Schmid1983,Furusaki1993,Giamarchi2003}
we can write down trajectories for six instantons between successive
minimal of the cosine potentials (our choice of resonance condition
means that processes involving $\theta_{-c}$ that change the charge
on the dot are less important than other terms). Furthermore, the
operator $\cos\bar{\phi}_{+c}$ in Eq. \eqref{eq:kondotunnelingappendix}
is exactly the operator\cite{Saleur1999} that creates a half-instanton
in the original $\cos\theta_{+c}/\sqrt{2}$ potential, and similarly
for the other two fields. In this sense, the perturbation (Coulomb
gas) expansion of the tunneling Hamiltonian Eq. \eqref{eq:kondotunnelingappendix}
is identical to the instanton expansion of Eq.~(\ref{eq:dualmap}),
so one may regard $t_{_{e}}$, $\tilde{t}_{_{c}}$ and $\tilde{t}_{_{s}}$
as the fugacities of instantons in the original problem. Again we
see why there must be special treatment of the $\theta_{-s}$ (or
in the dual problem $\bar{\phi}_{-s}$) field: there are only two
allowed spin states of the dot, so the time ordering in this field
must alternate between instantons and anti-instantons. Treating this
carefully\cite{Kane1992} requires the introduction of a new parameter
$K_{s}$ into the RG equations in Eqs. \eqref{eq:resonnatOurAll}.

Finally, we show the relation between our notation and that of the
two-channel Kondo model. As the dot has only two states, we can replace
all operators on the dot (in the low energy limit) by a single spin-half
operator $\vec{S}$. The standard notation for the Kondo Hamiltonian
is then 
\begin{equation}
H^{K}=\sum_{i=1,2}J_{1}\vec{S}\cdot\left(\Psi_{i\sigma}\vec{\sigma}_{\sigma\sigma'}\Psi_{i\sigma'}\right)+\sum_{i\ne j}J_{2}\vec{S}\cdot\left(\Psi_{i\sigma}\vec{\sigma}_{\sigma\sigma'}\Psi_{j\sigma'}\right)\label{eq:kondostandard}
\end{equation}
 Comparing with our terms, we therefore see that $J_{1}^{xy}=\tilde{t}_{_{s}}$,
$J_{2}^{z}=t_{_{e}}$ and $J_{2}^{xy}=\tilde{t}_{_{c}}$. The term
we neglect is $J_{1}^{z}$.

Looking at Eq. \eqref{eq:kondostandard}, we see that the $J_{1}$
terms are the traditional Kondo spin-exchange couplings to the two
leads, while the $J_{2}$ terms are those that allow for charge transport.
The two-channel Kondo fixed point is therefore the one where $J_{2}=0$,
i.e. only the Kondo couplings remain, and furthermore flow to strong
coupling. This is exactly the phase we call IC -- note that while
the II phase also has $J_{2}\rightarrow0$ under RG, in this case
$J_{1}$ also flows to weak coupling which is the decoupled dot fixed
point and not a Kondo-like one. On the other hand, if both $J_{1}$
and $J_{2}$ flow to strong coupling, this is the single channel Kondo
fixed point.\cite{Glazman1988,Ng1988}


\section{\label{sec:Analytical}Analytical analysis of the separatix line
between the CC and II phases at $1<g<2$}

In this appendix, we derive the limits of the phase boundary in Eq.~(\ref{eq:tcbound})
in the main text by giving an approximate analytic solution to the
flow equations in Eq.~\eqref{eq:resonnatOurAll}.

Without the quadratic terms $A=0$, the flow equation for $t$$_{_{e}}$
\eqref{eq:resonnatOur1} decouples from the rest and always flows
to weak coupling. However, the flow of $\tilde{t}_{_{c}}$ and $\tilde{t}_{_{s}}$
are less trivial since they both depend on $K$. To make progress,
we rearrange Eqs. (\ref{eq:resonnatOur2} - \ref{eq:kdependece})
as follows, 
\begin{eqnarray}
\frac{d\tilde{t}_{_{c}}}{dl}-a\tilde{t}_{_{c}} & = & -\frac{K_{s}g}{2}\tilde{t}_{_{c}},\label{eq:resonnatOur2-1}\\
\frac{d\tilde{t}_{_{s}}}{dl}-b\tilde{t}_{_{s}} & = & -\frac{K_{s}g}{2}\tilde{t}_{_{s}},\label{eq:resonnatOur3-1}
\end{eqnarray}
 where $a=1-\frac{1}{2g}$, and $b=1-\frac{g}{2}$.

Introducing new variables 
\[
X=\tilde{t}_{_{c}}e^{-al},\quad\quad Y=\tilde{t}_{_{s}}e^{-bl},
\]
 we find \begin{subequations} 
\begin{eqnarray}
\frac{dX}{dl} & = & -\frac{K_{s}g}{2}X,\label{eq:xfun}\\
\frac{dY}{dl} & = & -\frac{K_{s}g}{2}Y.\label{eq:yfun}
\end{eqnarray}
 \end{subequations} Dividing Eq. \eqref{eq:xfun} by Eq. \eqref{eq:yfun}
gives $dX/dY=X/Y$ which has the solution $X=cY$ for some constant
$c$. Hence transforming back to our original variables gives 
\begin{equation}
\tilde{t}_{_{s}}(l)=\tilde{t}_{_{c}}(l)e^{(b-a)l},\label{eq:tstcrela}
\end{equation}
 and our problem now reduces to solving two coupled differential equations
for $\tilde{t}_{_{c}}$ and $K_{s}$. The equation for the former
is given in Eq. \eqref{eq:resonnatOur2-1} while the latter now satisfies
\begin{equation}
\frac{dK_{s}}{dl}=-\left(\frac{1}{g}+g\alpha(l)\right)\tilde{t}_{_{c}}^{2}K_{s},\label{eq:kdependece-1}
\end{equation}
 where $\alpha(l)=\frac{\tilde{t}_{_{s}}^{2}(0)}{\tilde{t}_{_{c}}^{2}(0)}e^{2(b-a)l}$
which falls between $0$ and $\alpha_{\mathrm{max}}$ for $l>0$ as
long as $g>1$. In the case that the bare tunneling terms are identical,
$\alpha_{\mathrm{max}}=1$.

We are interested in the fixed point of the flow in Eqs. \eqref{eq:resonnatOur2-1}
and \eqref{eq:kdependece-1} in the limit $l\rightarrow\infty$. There
are two possibilities: either $\tilde{t}_{_{c}}\rightarrow\infty$
and $K_{s}\rightarrow0$ which is the conducting phase, or $\tilde{t}_{_{c}}\rightarrow0$
with $K_{s}\rightarrow K_{*}$. It is not difficult to conclude that
in the region of interest, $1<g<2$, one must have $K_{*}>0$. We
now understand the shape of the flow of the equations -- as $l$ grows,
$K_{s}$ decreases; if it decreases past a certain point then the
eventual fixed point is ultimately the conducting one. This invites
an easy approximation to allow us to approximately locate the phase
boundary: if we neglect the flow of $\alpha$, i.e. let $\alpha(l)=\alpha_{\mathrm{max}}$,
then the negative flow of $K$ will be faster than the true flow,
and we will underestimate the critical bare $\tilde{t}_{_{c}}$ needed
for the system to reach the strong coupling fixed point. On the other
hand, simply putting $\alpha=0$ will do the opposite, and overestimate
the phase boundary. Hence by solving the equations for constant $\alpha$
and finally substituting in the values $\alpha=0,\alpha_{\mathrm{max}}$
we find upper and lower bounds on the phase boundary line.

\begin{figure}
\includegraphics[width=1\columnwidth]{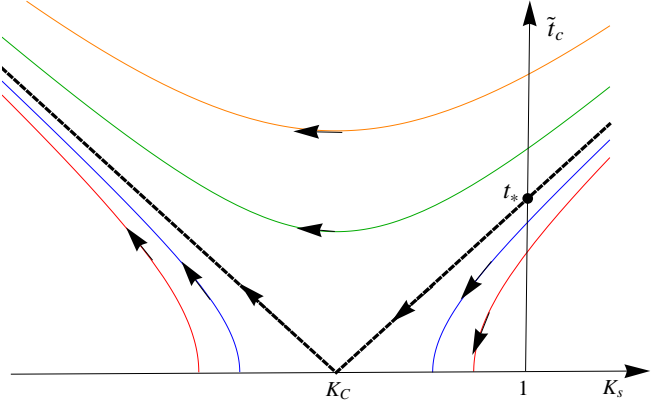} \caption{\label{fig:Flow}Renormalization-group flows for resonant tunneling
in TIs for $1<g<2$, given by Eq.~\eqref{eq:tcsolution}. The different
lines indicate different initial parameter $t_{0}$; the RG flows
always starts from $K_{s}=1$. The thick dashed line represents the
separatrix, which is of the Kosterlitz-Thouless universality class.
When $t_{0}$ is larger than $t_{*},$ the system flows to the conducting
phase, while for $t_{0}$ smaller than $t_{*}$, the system flows
to the insulating phase. This plot is made at $g=1.05$.}
\end{figure}

If $\alpha$ is a constant, we can divide Eq.~(\ref{eq:resonnatOur2-1})
by (\ref{eq:kdependece-1}) and integrate to obtain 
\begin{equation}
\tilde{t}_{_{c}}^{2}=t_{0}^{2}+\frac{\left(K_{s}-1\right)g^{2}}{1+\alpha g^{2}}-\frac{2g-1}{1+\alpha g^{2}}\ln K_{s}.\label{eq:tcsolution}
\end{equation}
 where $t_{0}=\tilde{t}_{_{c}}(l=0)$ is the bare value of the tunneling.
Fig.~\ref{fig:Flow} show a plot of this solution for various different
parameters $t_{0}$. From the plot, the strategy to find the critical
$t_{0}\equiv t_{*}$ is clear. If $\tilde{t}_{_{c}}^{2}$ remains
positive, then the eventual fixed point of the flow will be $K=0$,
$\tilde{t}_{_{c}}\rightarrow\infty$; while if $\tilde{t}_{_{c}}^{2}=0$
anywhere along the flow line, then this is the fixed point (a similar
situation occurs in the Kosterlitz-Thouless phase diagram).

By differentiating, we find that the minimum value of $\tilde{t}_{_{c}}^{2}$
occurs at a value 
\begin{equation}
K_{c}=\frac{2}{g}-\frac{1}{g^{2}}.\label{eq:Kc-1}
\end{equation}
 The critical $t_{*}$ is therefore given by the solution to the equation
$\tilde{t}_{_{c}}^{2}(K_{c})=0$; executing the calculation yields
\begin{equation}
t_{*}^{2}=\frac{\left(g-1\right)^{2}}{1+\alpha g^{2}}+\frac{2g-1}{1+\alpha g^{2}}\ln\left[1-\left(\frac{g-1}{g}\right)^{2}\right].\label{eq:crticaltvsg-1}
\end{equation}
 In the vicinity of $g=1+\varepsilon$ with $\varepsilon\ll1$, expanding
\eqref{eq:crticaltvsg-1} gives 
\begin{equation}
\frac{\varepsilon^{4}}{2(1+\alpha_{\mathrm{max}})}<t_{*}^{2}<\frac{\varepsilon^{4}}{2}.\label{eq:tcbound-1}
\end{equation}
 Assuming that all bare tunneling amplitudes are equal so $\alpha_{\textrm{max}}=1$,
this gives the condition in Eq. \eqref{eq:tcbound} quoted in the
main text. These bounds, along with a numerical determination of the
phase boundary are shown in Fig.~\ref{fig:CC-and-II}.

We now discuss briefly how the phase boundary is affected if the quadratic
terms in Eq. \eqref{eq:resonnatOurAll} are included, i.e. if we set
$A\ne0$. In the vicinity of $g=1$ where the linear terms may be
zero, it is not a priori obvious that one can ignore the quadratic
terms. However, the fact that the phase boundary line approaches zero
as $\varepsilon^{2}$ means that in the vicinity of the transition
line, such terms are of order $\varepsilon^{4}$ while the first term
behaves as $\varepsilon^{3}$. We therefore conclude that including
the quadratic terms does not significantly affect the phase boundary
-- this is also backed up by the numerical plot of the phase boundary
at $A=1$ in Fig.~\ref{fig:CC-and-II}.

\section{Analytical analysis of the RG flow}

\label{sec:RGflow}

In this appendix, we analyse the flow of Eqs.~\eqref{eq:resonnatOurAll}
as a function of $l$ in order to locate $l^{*}$ and $l_{K}$, corresponding
through the relation $l=\ln(\Lambda/T)$ to the minimum-conductance
temperature and Kondo temperature respectively. Our goal will be parametric
relations when the bare parameters are small, hence we will ignore
the quadratic terms (i.e. set $A=0$), which means that as usual,
the equation for $t_{_{e}}$ decouples, and we can ignore it. The
initial conditions are that $\tilde{t}_{_{c}}(l=0)=\tilde{t}_{_{s}}(l=0)=t_{0}$
and $K_{s}(l=0)=1$.

We begin by noticing that we can write the formal solution to Eq.~\eqref{eq:resonnatOur2}
as 
\begin{equation}
\tilde{t}_{_{c}}(l)=t_{0}\exp\left[(1-1/2g)l-\frac{g}{2}\int_{0}^{l}K_{s}(l')dl'\right].\label{eq:tcsoln}
\end{equation}
 By substituting this into Eq.~\eqref{eq:kdependece} and further
using \eqref{eq:tstcrela}, we find that 
\begin{multline}
\frac{1}{K_{s}}\frac{dK_{s}}{dl}=\\
-t_{0}^{2}\left[\frac{e^{(2-1/g)l}}{g}+ge^{(2-g)l}\right]\exp\left[-g\int_{0}^{l}K_{s}(l')dl'\right].\label{eq:Kexact}
\end{multline}
 Now initially, the flow of $K_{s}$ is close to one, so approximating
$K_{s}$ on the RHS of \eqref{eq:Kexact} by $1$, we can integrate
to obtain 
\begin{equation}
K_{s}(l)=\exp\left\{ -t_{0}^{2}\left[\frac{e^{(2-g-1/g)l}-1}{g(2-g-1/g)}+\frac{g(e^{(2-2g)l}-1)}{2-2g}\right]\right\} .\label{eq:Kapprox}
\end{equation}
 where we have additionally taken $g\ne1$. This expression is fine
until $K_{s}$ starts differing significantly from $1$. In particular,
we can use it until the scale $l^{*}$ defined by $K_{s}(l^{*})=K_{c}$
with $K_{c}$ given in Eq.~\eqref{eq:Kc-1}. This is exactly the
scale at which $\tilde{t}_{_{c}}(l)$ takes a minimum value, and is
given by the solution of 
\begin{equation}
\frac{e^{(2-g-1/g)l^{*}}-1}{g(2-g-1/g)}+\frac{g(e^{(2-2g)l^{*}}-1)}{2-2g}=\frac{\ln(g^{2}/(2g-1))}{t_{0}^{2}}.\label{eq:longlstar}
\end{equation}
 It is worth pointing out that if $g>1$, then the two terms on the
left hand side are exponentially decaying, and therefore this equation
may not have a solution with $l^{*}>0$ if $t_{0}$ is too small.
This is an alternative way of looking at the phase transition.

Limiting ourselves to parameters where there is a solution, writing
$g=1+\varepsilon$ and expanding in small $\varepsilon$ gives expression
\eqref{eq:Tstar1} in the main text. Assuming $g<1$, the second term
on the LHS of \eqref{eq:longlstar} is exponentially growing and therefore
dominant; by taking the leading behavior and ignoring prefactors gives
expression \eqref{eq:Tstar2} in the main text. Substituting this
value of $l^{*}$ into \eqref{eq:tcsoln} (and again approximating
$K_{s}=1$ on the RHS) gives Eq.~\eqref{eq:Gmin} in the main text.

Now, for $l>l^{*}$, we can no longer approximate $K_{s}$ as $1$.
However for $g<1$ (or more formally $(1-g)/t_{0}\gg1$), we can make
progress. In this limit, due to the exponentially increasing term
on the RHS of \eqref{eq:Kapprox}, we see that $K_{s}$ drops very
rapidly to become close to zero. Hence to leading order, we can say
that 
\begin{equation}
\int_{0}^{l>l^{*}}K_{s}(l')dl'\approx l^{*},
\end{equation}
 and hence from \eqref{eq:tcsoln} we obtain 
\begin{equation}
\tilde{t}_{_{c}}(l>l^{*})=t_{0}\exp\left[(1-1/2g)l-gl^{*}/2\right].
\end{equation}
 Solving this to find the scale where $t_{_{c}}(l_{K})=1$ gives expression
\eqref{eq:Tkondo} in the main text, while the flow of $\tilde{t}_{_{c}}$
between $l^{*}$ and $l_{K}$ gives the intermediate-scale power-law
quoted in the main text.

For $g>1$, the approximations leading to the results about $T_{K}$
no longer are valid -- the lack of an exponentially increasing factor
in Eq.~\eqref{eq:Kapprox} means that the deviation of $K_{s}$ from
one in the RHS of Eq.~\eqref{eq:Kexact} is crucial. In fact, looking
at numerical results, one can see that in this case, the decrease
in $K_{s}$ from $1$ to $0$ happens exactly over the same energy
region as $\tilde{t}_{_{c}}$ flows to strong coupling. This intertwining
of energy scales means that there is no great separation in scale
of $T^{*}$ and $T_{K}$, and hence the intermediate regime in the
pictures may be rather narrow. We therefore say no more about this
region here.

 \bibliographystyle{apsrev}

\end{document}